\documentclass{ieeeaccess}
\usepackage{cite}
\usepackage{amsmath,amssymb,amsfonts}
\usepackage{algorithmic}
\usepackage{graphicx}
\usepackage{textcomp}
\usepackage{tikz}
\usetikzlibrary{shapes.geometric, arrows,automata,positioning,shapes.multipart}
\usepackage{tabularx,colortbl}

\definecolor{burntorange}{rgb}{0.8, 0.28, 0.0}
\definecolor{myGreen}{rgb}{0.0, 0.5, 0.0}
\definecolor{amber}{rgb}{0.8, 0.28, 0.0}
\definecolor{ceruleanblue}{rgb}{0.16, 0.28, 0.75}

\NewSpotColorSpace{PANTONE}
\AddSpotColor{PANTONE} {PANTONE3015C} {PANTONE\SpotSpace 3015\SpotSpace C} {1 0.3 0 0.2}
\SetPageColorSpace{PANTONE}%

\usepackage{enumitem}
\usepackage{overpic}
\usepackage[crop=pdfcrop]{pstool}
\usepackage{color}
\usepackage[font=footnotesize]{subcaption}
\usepackage{mathdots}
\usepackage{color}
\usepackage{mathtools, cuted}
\usepackage{bbm}
\usepackage{dblfloatfix}    % To enable figures at the bottom of page
\usepackage{bbm}
\usepackage{array}
\usepackage{float}
\usepackage{srcltx}
\usepackage{color}
\usepackage{mathdots}
\usepackage[colorlinks=true, allcolors=blue]{hyperref}
\usepackage[export]{adjustbox}
\usepackage[normalem]{ulem} %to strike the words
\usepackage{color}
\usepackage{float}
\usepackage{mathtools, cuted}
\usepackage[makeroom]{cancel}
\usepackage{relsize}
\usepackage{bbm}
\usepackage{ulem}
\usepackage{flushend}
\usepackage{subcaption}

\newcommand{\bbmatrix}{\begin{bmatrix}}
\newcommand{\ebmatrix}{\end{bmatrix}}

\newcommand{\E}{\mathbf{E}}

\renewcommand{\H}{\mathbf{H}}

\newcommand{\NN}{\mathbb{\bar N}}
\newcommand{\J}{\mathbf{J}}
\newcommand{\K}{\mathbf{K}}

\renewcommand{\r}{\mathbf{r}}

\newcommand{\rb}{\boldsymbol{\r}}
\newcommand{\rbb}{\mathbbm{r}}
\renewcommand{\L}{\mathbf{\mathcal{L}}}
\newcommand{\R}{\mathbf{\mathcal{R} }}
\newcommand{\Nh}{\mathbf{\hat n}}
\newcommand{\0}{\varnothing}
\newcommand{\I}{\mathbb{I}}

\newcommand{\Cv}{\mathbb{C}}
\newcommand{\Qv}{\mathbb{\bar Q}}
\newcommand{\Av}{\mathbb{A}}
\newcommand{\Bv}{\mathbb{B}}
\newcommand{\Pv}{\overline{\mathbb{P}}}

\newcommand{\Sv}{\mathbb{S}}
\newcommand{\Fv}{\mathbb{F}}
\newcommand{\Tv}{\mathbb{T}}

\newcommand{\Rv}{\mathbb{R}}
\newcommand{\Xv}{\mathbb{X}}

\newcommand{\Lv}{\mathbb{L}}
\newcommand{\Ev}{\mathbb{E}}

\newcommand{\Dv}{\mathbb{\bar D}}
\newcommand{\Gv}{\mathbb{\bar G}}
\newcommand{\Jv}{\mathbb{J}}
\newcommand{\Hv}{\mathbb{H}}
\newcommand{\Kv}{\mathbb{K}}

\newcommand{\Nv}{\overline{\mathbb{N}}}

\def\BibTeX{{\rm B\kern-.05em{\sc i\kern-.025em b}\kern-.08em
    T\kern-.1667em\lower.7ex\hbox{E}\kern-.125emX}}
\begin{document}

\history{MANUSCRIPT DRAFT}
\doi{}

\title{Surface Susceptibility Synthesis of Metasurface Skins/Holograms for Electromagnetic Camouflage/Illusions}

\author{\uppercase{Tom J. Smy} and
\uppercase{Shulabh Gupta},
\IEEEmembership{Senior Member, IEEE}}
\address{Department of Electronics (DoE), Carleton University, Ottawa, Ontario, Canada, K1S 5B6}
\tfootnote{The authors acknowledge funding from the Department of National Defence's Innovation for Defence Excellence and Security (IDEaS) Program in support of this work.}

\markboth
{Smy \headeretal: Surface Susceptibility Synthesis of Metasurface Skins/Holograms}
{Smy \headeretal: Surface Susceptibility Synthesis of Metasurface Skins/Holograms}

\corresp{Corresponding author: Tom. J. Smy (e-mail: tjsmy@cunet.carleton.ca).}

\begin{abstract}
A systematic numerical framework based on Integral Equations and Generalized Sheet Transition Conditions (IE-GSTCs) is presented in 2D to synthesize closed metasurface holograms and skins for creating electromagnetic illusions of specified objects and as a special case, to camouflaging them against their backgrounds. The versatile hologram surface is modeled using a zero-thickness sheet model of a generalized metasurface expressed in terms of its surface susceptibilities, which is further integrated into the GSTCs and the IE current-field propagation operators. To estimate the effectiveness of the illusions, the notion of a scene constructed by an observer is developed from first principles and a simple mathematical model, referred to as a Structured Field Observation (SFO), based on spatial Fourier transform is proposed. Using numerical examples, it is shown that to recreate the reference desired fields everywhere in space using a closed metasurface hologram/skin, an internal illumination must be applied inside the hologram, in addition to the applied external illumination fields. Finally, several numerical examples are presented for simple, angle-dependent and dynamic illusions. Finally, a dynamic camouflaged region of space, which can freely move inside a given complex scene without being detected by the observer is demonstrated.
\end{abstract}

\begin{keywords}
Electromagnetic Metasurfaces, Metasurface Holograms, Effective Surface Susceptibilities, Boundary Element Method (BEM), Generalized Sheet Transition Conditions (GSTCs), Method of Moments (MoM), Field Scattering, Electromagnetic Illusions, Electromagnetic Camouflage.
\end{keywords}

\titlepgskip=-15pt

\maketitle

\section{Introduction}

Electromagnetic (EM) invisibility has gathered an immense interest in the past two decades as a result of a rapid development in the general area of electromagnetic metamaterials. Metamaterials led the way to the realization of  cloaking devices based on transformation optics to enclose an object which ``bends'' the light around the object so that there are no scattered fields that reach the observer, i.e. minimizing the Radar Cross Section (RCS) of the object. The object appears invisible to the observer. However, such cloaks are based on volumetric shells with extremely challenging nonuniform, anisotropic and active material requirements  \cite{CloakingReview, YAN2009261, ALITALO200922}, which has led to investigation of alternate routes to make an object invisible or undetectable.

An alternative to electromagnetic cloaking is an \textit{Electromagnetic Camouflage} which results in an \textit{effective} electromagnetic invisibility making an object hard to detect. The object inherits the scattering property of the background it is in, and ``blends" into the background. Consequently, an observer cannot distinguish the fields scattered off from the object and the background. While there is an apparent lack of electromagnetic cloaks in nature, camouflage appears in nature in myriad of ways ranging from static conditions to dynamic camouflage of cephalopods. This suggests that camouflage is a better practical alternative compared to cloaks in the evolutionary sense. In spite of sharing the same goal, cloaking and camouflaging have fundamentally different operating mechanism: where cloaking aims at minimizing the RCS, the camouflage has a non-zero RCS which is engineered to match its background. 

Electromagnetic camouflage is thus a form of illusion which is similar to what is produced using \textit{Holograms}. Holograms are well-known in optics where the spatial (and possibly temporal) information of an arbitrary object is encoded onto the surface (typically photographic plates) \cite{Goodman_Fourier_Optics, Saleh_Teich_FP}. This is a two step process, where the information about scattering properties of an object of interest is first recorded using a given reference beam (i.e. incident fields) and modulated onto a given surface. Once the information is recorded, the encoded surface, when illuminated with a reconstructing beam (i.e. illumination fields)\footnote{In typical holograms the illumination fields are the same as the incident fields used in the first stage of information encoding on the plates. However, a distinction is made here between the two, where the information maybe encoded in such a way, that the object may still be recreated with a different illumination field.}, projects an illusion of the object. With increasing sophistication of encoding capability, more complex illusions can naturally be created. Typically, holograms are encoded with object information that one wants to project as an illusion. However if the object itself is a hologram or is enclosed inside a holographic \textit{skin} which is encoded with the scattering information of the entire environment including the background, the object is camouflaged against the background. Camouflaging thus can be seen as a special case of electromagnetic holograms.

Creating holograms naturally demands a versatile and a flexible surface that can be engineered to project myriad of illusions, including camouflage. To realize such holograms, \textit{Electromagnetic Metasurfaces} represent a powerful platform due to their complete control over the scattered fields with respect to both complex amplitude and polarization. They are 2D arrays of sub-wavelength resonating particles, where control of the spatial distribution and EM properties of the individual particles allows the scattered fields to be engineered with unprecedented control of both reflection and transmission, and with complete polarization control \cite{meta2,MS_review_Yu}. Consequently, these surfaces have been used to create \textit{metasurface holograms}, due to their advanced information encoding capability \cite{MSHologram_Review, MShologram_Review2}. 

To enable a general treatment of the problem, practical metasurfaces can conveniently be modeled as zero thickness sheets characterized using frequency dependent electromagnetic surface susceptibility tensors $\bar{\bar{\chi}}(\omega)$ \cite{Chi_Review, MS_Synthesis, Chi_extraction_Macrodmodel, TBC_vs_GSTC_Caloz}. The EM fields around the metasurface then can be described using Generalized Sheet Transition Conditions (GSTCs) \cite{KuesterGSTC}. The spatial distribution of surface susceptibilities of the metasurface $\bar{\bar{\chi}}(\mathbf{r})$ dictates the scattered (and thus total) fields produced by the metasurface when illuminated by an incident field. Therefore, the key design objective in creating metasurface based illusions (and object camouflage) is to synthesize the spatially varying surface susceptibilities, $\bar{\bar{\chi}}(\mathbf{r})$. 

A systematic description of \textit{open} metasurface holograms based on GSTCs and surface susceptibility description was recently presented in \cite{smy2020surface} where various classifications and rigorous procedures were defined to design and synthesize these metasurfaces for achieving a desired EM illusion. Due to the open nature of the metasurface, the illusion could only be created in a half-space. In this work, the surface susceptibility synthesis of metasurface holograms is extended to \textit{closed metasurfaces} recreating the desired fields, so that an observer can move around the metasurface and perceive the illusion from different directions. Moreover, in defining an {\it illusion} we extend this field recreation and place it in a {\it scene} consisting of an incident field and other objects not part of the intrinsic illusion, as opposed to a standalone hologram used in \cite{smy2020surface}. This also enables camouflaging an object against its background. Transition from an open to a closed metasurface also has important implications for the illumination fields. As will be shown later, an external illumination of the closed metasurface is not sufficient to fully reconstruct the desired reference fields everywhere in the region. Consequently, an internal illumination in addition to the external illumination is proposed to be a potential solution and it will be shown that a proper choice of internal illumination enables a complete reconstruction of the fields everywhere in space.

Many metasurface synthesis and analysis problems using surface susceptibilities have been reported in the literature, where for planar surfaces, metasurface susceptibilities can be analytically computed, for instance \cite{MS_Synthesis, Caloz_EM_inversion}. On the other hand, metasurface analysis typically involves integrating GSTCs into bulk Maxwell's equations using a variety of standard numerical techniques based on Finite-Difference and Finite Element methods \cite{Caloz_MS_Siijm, Caloz_Spectral, Smy_Metasurface_Space_Time}, and Integral-Equation (IE) techniques \cite{stewart2019scattering, FE_BEM_Impedance, Caloz_MS_IE, AppBEMEM, Smy_EuCap_BEM_2020, Caloz_EM_inversion}. Given that the field scattering from a metasurface hologram may need to be evaluated for electrically large domains, IE-GSTC methods are a computationally efficient choice and, as will be shown later and in \cite{smy2020surface}, are well suited for analysis of both closed and open metasurface holograms.

Given this context, a general methodology of synthesizing and designing closed metasurface holograms, which may be located within a complex environment with variety of objects and backgrounds, is presented in this work using the IE-GSTC method. The 2D IE-GSTC based numerical platform is further developed to synthesize metasurface susceptibilities with an integrated approach, where the desired fields, specified anywhere in space and not necessarily at the metasurface, are generated using a system level description and fed-back into the metasurface design procedure. A novel proposal of utilizing both external and internal illumination fields inside closed metasurface holograms is presented to rigorously reconstruct the desired fields everywhere in space. Next, a simple but powerful method is proposed to model the observer and how it measures the total and scattered fields to construct a given scene of the environment. Finally, using various numerical examples, several metasurface holograms are synthesized to project complex angle dependent illusions and dynamic illusions where the surrounding objects may be moving with respect to the hologram. An example of an object camouflage is further presented, where a closed metasurface is navigating through a complex environment without being detected by the observer.

The paper is structured as follows. Sec.~II presents the general problem of illusion formation and object camouflage using metasurfaces, and how an observer constructs the image of the environment. The general goal of the metasurface synthesis is formulated and important aspects related to illumination fields are described. Sec.~III presents the IE-GSTC architecture, and presents a systematic way of designing an illusion including the object camouflage as its special case. A simple and elegant technique to model the observer and how it constructs the scene is presented in Sec.~IV. Several numerical demonstrations to illustrate the presented metasurface synthesis approach is given in Sec.~V, for cases of complex merged illusions and that in dynamic environments. Finally, a summary and concluding discussions are provided in Sec.~VI.

%A methodology for designing  holographic illusions using open metasurfaces based on an integral equation infrastructure was presented in \cite{smy2020surface} (afterwords referred to as Smy2020). In this paper we will differentiate between a holographic image and an illusion. A hologram is a stand-alone recreation of the field distribution for an object subject to a incident field. The final hologram consists of a encoded surface (in Smy2020 a synthesized metasurface) and a illuminating field. In defining an {\it illusion} we extend this field recreation and place it in a {\it scene} consisting of an incident field and other objects not part of the intrinsic illusion.

\section{Problem Formulation}\label{Sec:II}

\subsection{The Observer \& Scenes}

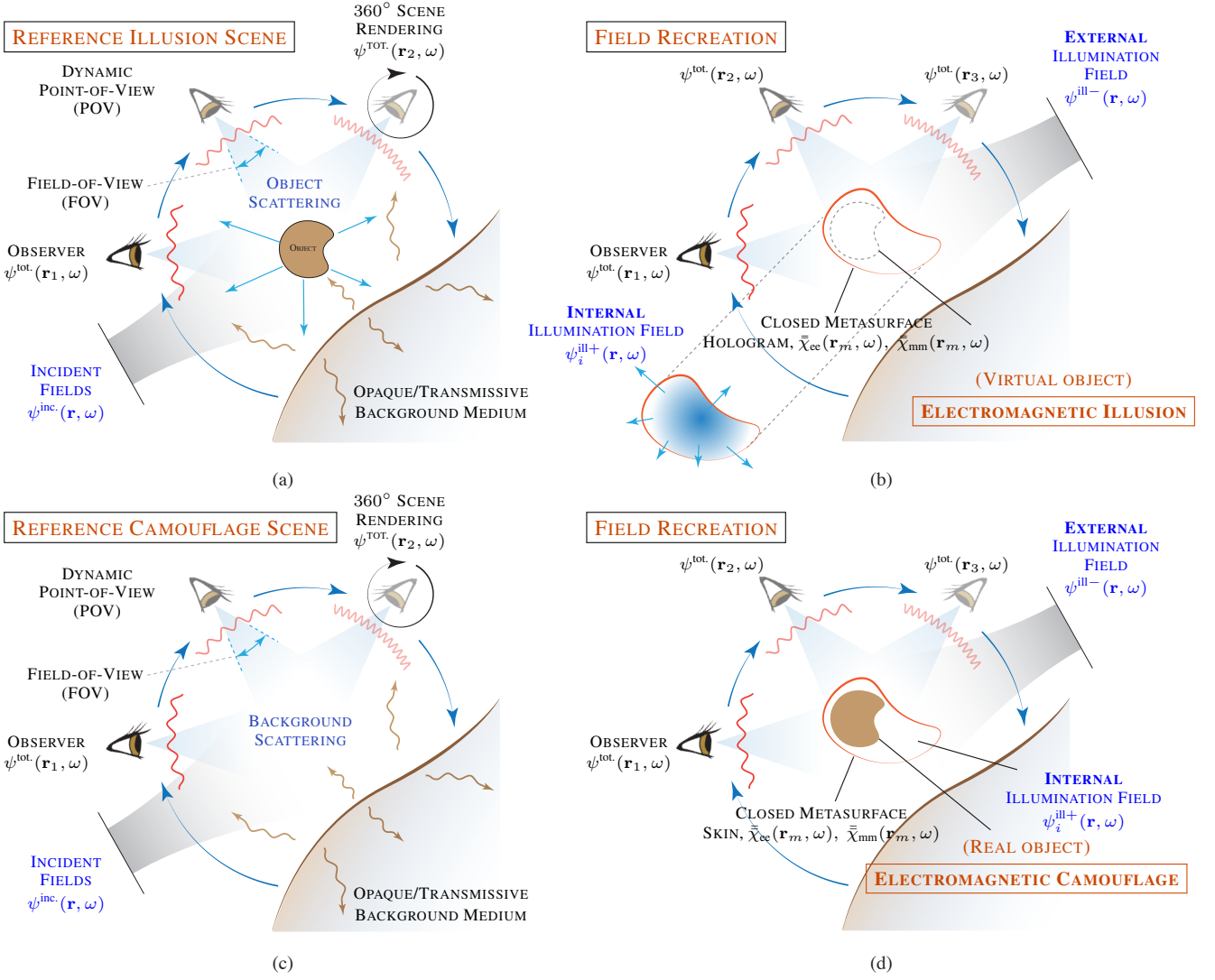
\begin{figure*}[h]
\centering
	\begin{overpic}[grid=false, scale = 0.22]{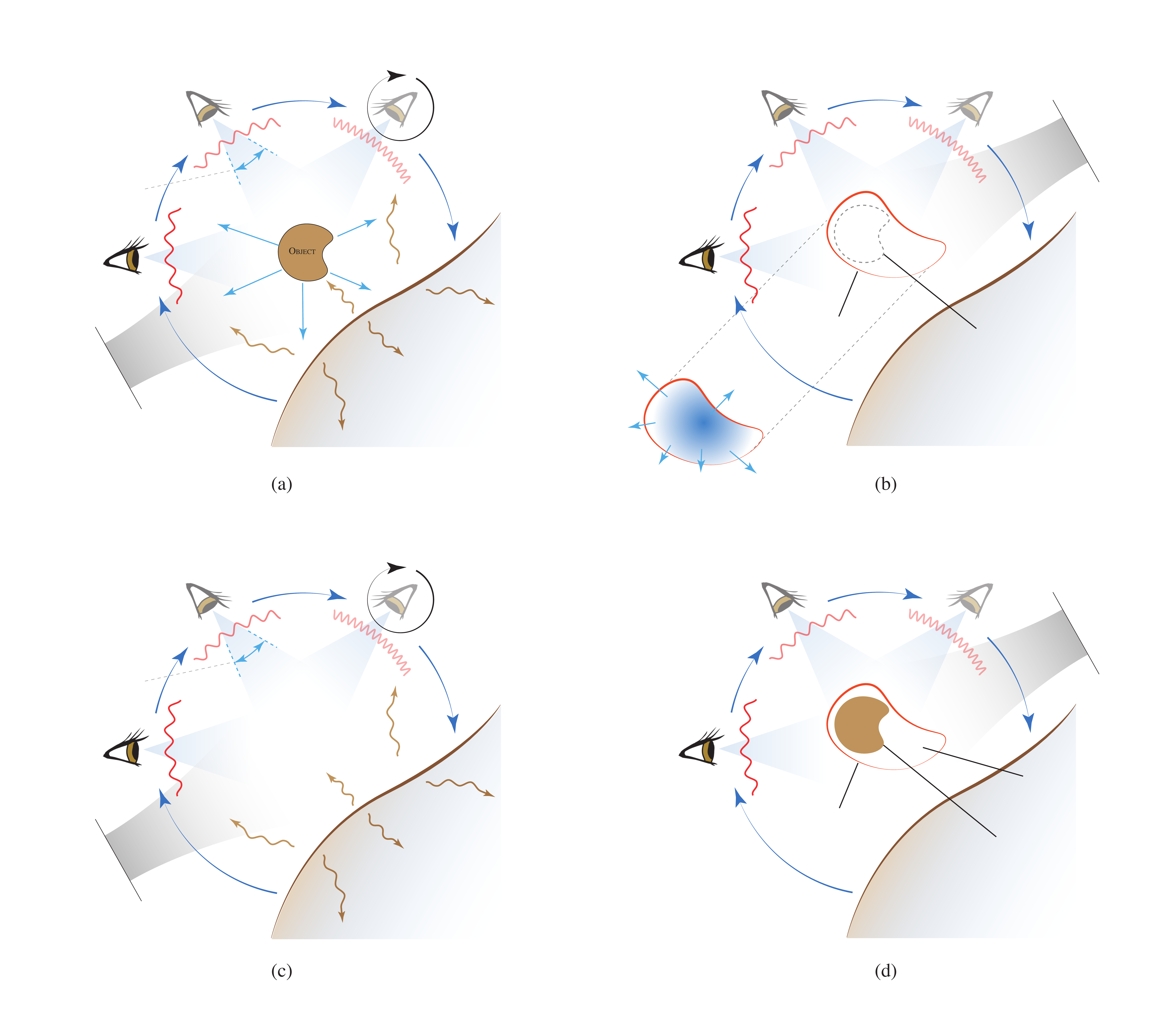}
		\put(0,62){\scriptsize{\shortstack{\textsc{Observer}\\ $\psi^\text{tot.}(\mathbf{r}_1,\omega)$}}}
		\put(0,20){\scriptsize{\shortstack{\textsc{Observer}\\ $\psi^\text{tot.}(\mathbf{r}_1,\omega)$}}}
		\put(50,20){\scriptsize{\shortstack{\textsc{Observer}\\ $\psi^\text{tot.}(\mathbf{r}_1,\omega)$}}}
		\put(50,62){\scriptsize{\shortstack{\textsc{Observer}\\ $\psi^\text{tot.}(\mathbf{r}_1,\omega)$}}}
		\put(58, 79){\scriptsize{$\psi^\text{tot.}(\mathbf{r}_2,\omega)$}}
		\put(58, 37){\scriptsize{$\psi^\text{tot.}(\mathbf{r}_2,\omega)$}}
		\put(79,79){\scriptsize{$\psi^\text{tot.}(\mathbf{r}_3,\omega)$}}
		\put(79,37){\scriptsize{$\psi^\text{tot.}(\mathbf{r}_3,\omega)$}}
		\put(0, 82){\small\boxed{\color{amber}\textsc{Reference Illusion Scene}}}
		\put(0, 40){\small\boxed{\color{amber}\textsc{Reference Camouflage Scene}}}
		\put(50, 82){\small\boxed{\color{amber}\textsc{Field Recreation}}}
		\put(50, 40){\small\boxed{\color{amber}\textsc{Field Recreation}}}
		\put(2, 50){\scriptsize\color{blue}\shortstack{\textsc{Incident}\\ \textsc{Fields}\\ $\psi^\text{inc.}(\mathbf{r},\omega)$}}
		\put(2, 8){\scriptsize\color{blue}\shortstack{\textsc{Incident}\\ \textsc{Fields}\\ $\psi^\text{inc.}(\mathbf{r},\omega)$}}
		\put(21, 68){\scriptsize\color{ceruleanblue}\shortstack{\textsc{Object}\\\textsc{Scattering}}}
		\put(21, 22){\scriptsize\color{ceruleanblue}\shortstack{\textsc{Background}\\\textsc{Scattering}}}	
		\put(30, 50){\scriptsize\shortstack{\textsc{Opaque/Transmissive} \\ \textsc{Background Medium}}}
		\put(30, 7){\scriptsize\shortstack{\textsc{Opaque/Transmissive} \\ \textsc{Background Medium}}}

		\put(90, 77){\scriptsize\color{blue}\shortstack{\textsc{\textbf{External}}\\\textsc{Illumination}\\ \textsc{Field}\\ $\psi^{\text{ill}-}(\mathbf{r},	\omega)$}}
		\put(90, 35){\scriptsize\color{blue}\shortstack{\textsc{\textbf{External}}\\\textsc{Illumination}\\ \textsc{Field}\\ $\psi^{\text{ill}-}(\mathbf{r},	\omega)$}}
		\put(45, 55){\scriptsize\color{blue}\shortstack{\textsc{\textbf{Internal}}\\\textsc{Illumination Field}\\ $\psi_i^{\text{ill}+}(\mathbf{r},\omega)$}}
		\put(86, 15){\scriptsize\color{blue}\shortstack{\textsc{\textbf{Internal}}\\\textsc{Illumination Field}\\ $\psi_i^{\text{ill}+}(\mathbf{r},\omega)$}}
		\put(60, 56){\scriptsize\shortstack{\textsc{Closed Metasurface}\\\textsc{Hologram}, $\bar{\bar{\chi}}_\text{ee}(\r_m, \omega),~\bar{\bar{\chi}}_\text{mm}(\r_m, \omega)$}}
		\put(60, 14){\scriptsize\shortstack{\textsc{Closed Metasurface}\\\textsc{Skin}, $\bar{\bar{\chi}}_\text{ee}(\r_m, \omega),~\bar{\bar{\chi}}_\text{mm}(\r_m, \omega)$}}
		\put(78, 50){\footnotesize\color{amber}\shortstack{ \textsc{(Virtual object)} \\ \boxed{\textsc{\textbf{Electromagnetic Illusion}}}}}
		\put(74, 10){\footnotesize\color{amber}\shortstack{ \textsc{(Real object)} \\ \boxed{\textsc{\textbf{Electromagnetic Camouflage}}}}}
		\put(30, 81){\scriptsize\textsc{\shortstack{$360^\circ$ Scene\\ Rendering \\ $\psi^\text{tot.}(\mathbf{r}_2,\omega)$}}}
		\put(30, 39){\scriptsize\textsc{\shortstack{$360^\circ$ Scene\\ Rendering \\ $\psi^\text{tot.}(\mathbf{r}_2,\omega)$}}}
		\put(2, 26){\scriptsize\textsc{\shortstack{Field-of-View\\ (FOV)}}}
		\put(3, 33){\scriptsize\textsc{\shortstack{Dynamic \\ Point-of-View\\ (POV)}}}
		\put(2, 68){\scriptsize\textsc{\shortstack{Field-of-View\\ (FOV)}}}
		\put(3, 76){\scriptsize\textsc{\shortstack{Dynamic \\ Point-of-View\\ (POV)}}}
	\end{overpic}
\caption{Illustration of using a closed metasurface shield to produce electromagnetic illusion or camouflaging an object. a) Reference scene to determine the desired scattered fields from an object of interest in the presence of a background. b) Introducing a metasurface as a hologram to create an illusion of the object (with the object removed). c) Reference scene for camouflaging where the observer records the scene without the object. d) Introducing a metasurface as a skin to enclose an object and camouflage it against the background.}\label{Fig:Illusion_Cases}
\end{figure*}

Consider a field scattering problem illustrated in Fig.~\ref{Fig:Illusion_Cases}(a), where a specified incident wave, $\psi^\text{inc.}(\mathbf{r},\omega)$ is illuminating an object of interest. The object may be placed in a certain environment or a \textit{Scene}, such as in front of a background which could be either a textured reflective/opaque surface or with partial reflection or transmission i.e. semi-transparent background. The incident field strikes different structures in the environment and various scattered fields from the object and the background are generated. Within this scene an \textit{Observer} can be placed in a location that views the object from a point-of-view (POV) over a prescribed field-of-view (FOV). We will refer to the Observer {\it rendering} the scene -- by which we mean the scene is scanned over the FOV and the intensity of the field determined as a function of the scanning angle; creating an image of the scene. The observer thus detects/measures the total fields $\psi^\text{tot.}(\mathbf{r}_1,\omega)$, in general, which can be decomposed into the incident field and the scattered fields. An intelligent observer can now identify the presence and features of the target object in that scene by processing the information contained in the total fields, in comparison to that of an isolated background. Let us call this scene as the \textit{Reference Scene}.

If the object is illuminated from the left and the Observer is placed on the left (a front-lit configuration) then the scattered fields traveling towards the Observer will be detected and the incident field will not, as it is traveling past and away from the Observer. On the other hand if the Observer is placed on the right side of the object the fields detected will be the total fields as both the incident and scattered fields are traveling towards the Observer [such as in $\psi^\text{tot.}(\mathbf{r}_2,\omega)$]. Due to this distinction the Observer needs to be characterized in terms of the \textit{directionality} and structure of the fields in the region nearby the Observer. For example, in the illustration of Fig.~\ref{Fig:Illusion_Cases}(a), there is an incident field present in the region to the left of the background. Both the object and the background surface produce scattered fields which in turn will scatter off each other. The net result of this is a complicated field pattern created by the interference of source field and the scattered field components. An observer placed within the region of interest will detect the EM waves propagating toward its POV within a FOV. If the POV is swept over a range of angles (e.g. by $360^\circ$) defined by the FOV, the scene will be rendered revealing the scattered object, reflected images, shadows on the background and complicated multipath effects, for instance.

\subsection{Principle of Illusion \& Camouflage}

We are now interested in artificially engineering the scene so that the observer either perceives a different object in the presence of the original object, or the same object in the absence of the original object. They correspond to the following two different, but closely related physical effects:

\begin{enumerate}[leftmargin=*]
\item \textit{Electromagnetic Illusion using a Metasurface Hologram:} Let us remove the object of interest from the scene of Fig.~\ref{Fig:Illusion_Cases}(a), and replace with an artificial closed surface at $\r = \r_m$ which is completely encompassing the object, as shown in Fig.~\ref{Fig:Illusion_Cases}(b). The objective is to engineer this surface so that the new scattered (and thus total) fields are identical to the reference scene at the observer location, when the object of interest was present. Since the observer measures the same fields, it falsely perceives the closed surface as a real object, while in reality, it is a virtual object, i.e. \textit{an electromagnetic illusion}. Such a closed surface, as will be shown later and throughout the paper, will be realized using an electromagnetic metasurface, and will now be referred to as a \textit{Metasurface Hologram}.

\item \textit{Electromagnetic Camouflage using a Metasurface Skin:} Now consider an alternate scenario where the object is first removed from the reference scene as shown in Fig.~\ref{Fig:Illusion_Cases}(c), so that the observer measures the fields corresponding to the background only in the presence of other possible scattering objects. We now introduce the object into the scene. Naturally, the introduction of this new object into the blank reference scene will perturb the fields. However, if the object is enclosed inside a closed engineered surface at $\r = \r_m$, as shown in Fig.~\ref{Fig:Illusion_Cases}(d), such that the newly generated fields are identical to the reference scene of Fig.~\ref{Fig:Illusion_Cases}(c), the observer will still perceive the background. Therefore, the object while being physically present in the scene is still undetected by the observer, where it has effectively blended into the background, i.e. \textit{an electromagnetic camouflage}. Such a closed surface, as will be shown later, will be realized using an electromagnetic metasurface, and will now be referred to as a \textit{Metasurface Skin}.

\end{enumerate}

Electromagnetic illusion and Camouflage, thus represent two practically important phenomena where the wave engineering capabilities of the metasurface becomes crucially useful. In spite of their apparent physical difference, there is a fundamental similarity between the two: while in the case of an electromagnetic illusion, the metasurface hologram mimics an object, the metasurface skin mimics the background in the case of an electromagnetic camouflage. In both cases, metasurface projects false information to the observer. Thus, the objective here now is to synthesize these metasurfaces enabling these operations. We will discuss the methods and procedures pertaining to a metasurface hologram only, while remembering that it is closely related to the camouflage operation.

\subsection{Metasurface Description}\label{Sec:II^-B}

An electromagnetic metasurface as remarked in the introduction, can be rigorously described using a zero thickness sheet model, using Generalized Sheet Transition Conditions (GSTCs) with specific electric and magnetic surface susceptibility densities $\bar{\bar{\chi}}_\text{ee}(\r_m, \omega)$ and $\bar{\bar{\chi}}_\text{mm}(\r_m, \omega)$. The problem of creating electromagnetic illusions and camouflage thus becomes the problem of determining the surface susceptibilities of closed metasurface holograms and skins. This formulation captures the general wave transformation capability of physical EM metasurfaces by expressing them as mathematical space discontinuities of zero thickness\cite{IdemenDiscont,GSTC_Holloway, KuesterGSTC}. The  GSTCs relate the tangential EM fields around the metasurface to the tangential and normal surface polarization response, rigorously modeling the EM interaction with the metasurface capturing the field transformation capabilities via 36 variables inside the susceptibility tensors. In this paper for simplicity we will limit the analysis to tangential terms and assume scalar susceptibilities and in the frequency domain we have for the GSTCs: \cite{smy2020surface,Chi_Review}
\begin{subequations}\label{Eq:GSTC}
\begin{align}
	\Nh \times \Delta \E_T &= -j\omega\mu_0 (\epsilon \chi_\text{mm} \H_{T,av} + \chi_\text{me} \sqrt{\epsilon/\mu}\; \E_{T,av})\\
	\Nh \times \Delta \H_T  &= j\omega (\epsilon \chi_\text{ee} \E_{T,av} + \chi_\text{em} \sqrt{\mu \epsilon}\; \H_{T,av}),
\end{align}
\end{subequations}
where $\Delta \psi_T = \psi(\r_{m+}) - \psi(\r_{m-})$, and $\psi_\text{av} = \{\psi(\r_{m+}) + \psi(\r_{m-})\}/2$, are expressed in terms of total fields just before and after the metasurface (this implies for a closed object that $^-$ indicates the region or field external to the object and $^+$ internal quantities). These equations can be incorporated into the IE infrastructure \cite{stewart2019scattering} to provide a complete simulation environment for creating metasurface based illusion and camouflage as will be described in following sections.

\subsection{The Illumination Fields}

In both cases of an electromagnetic illusion and camouflage, the closed metasurface may be excited with an external (to the metasurface) illumination field $\psi^{\text{ill}-}(\mathbf{r},\omega)$. This external illumination could either be the same as the incident field of the reference scene i.e. $\psi^{\text{ill}-}(\mathbf{r},\omega) = \psi^\text{inc.}(\mathbf{r},\omega)$ or be an entirely independent field, as is the case in Fig.~\ref{Fig:Illusion_Cases}(b). This figure shows the two possible sources of illumination external and internal. Initially we will consider the case of external illumination only as it would appear to a simple and obvious choice.

This illumination configuration however, presents a serious problem for the metasurface synthesis in order to recreate the fields of the reference scene, as the far side of the metasurface is in shadow and has effectively no field incident on it. However, for almost all illusions, this region of the metasurface will need to produce a scattered field (even if producing an illusory shadow, a scattered field needs to be produced to ``shape'' the shadow to match the recreated object). As we will see the synthesis process presented in Sec.~\ref{Sec:Arch}, will create the required susceptibilities to produce these scattered fields from the small incident fields present, however, these susceptibilities will be active, numerically difficult to handle and physically difficult to implement.

A possible solution to this issue is to use \textit{only} an internal illumination (as shown in Fig.~\ref{Fig:Illusion_Cases}b), where $\psi^{\text{ill}+}(\mathbf{r},\omega)$ is present inside the closed metasurface region. For this case we can achieve a uniform illumination of a transparent metasurface that modulates the internal source to produce the outgoing scattered fields recreating the illusion. Although this configuration allows for the synthesis of surface susceptibilities to recreate the scattered fields of the object, it can not recreate with any generality the previously present incident field used in the reference scene. This is an issue as from many view points the observer will register the effect of not only the scattered fields from the object but also the original incident field. 

Therefore, in the general case, both external and internal illumination will be required and it will be shown that such a configuration can optimally recreate the reference scene. Although the external illumination could be different from the incident field (with implications on the effectiveness of the illusion) the natural choice is to use an external illumination that is identical to the original incident field. For this case the entire metasurface is well illuminated (internally), and only has to recreate the original scattered fields propagating out from the object. This procedure allows for passive and well-characterized susceptibilities to be synthesized which have both reflective and transmissive aspects, as will be demonstrated later.

\section{Illusion Design Architecture}\label{Sec:Arch}

The primary requirements for the design of an illusion system are threefold: 1) the field distribution of the reference scene to be recreated (the object within its environment), 2) the specification of the illusion illumination, and 3) the synthesis of the metasurface used to create the illusion. To determine the fields present in the scene to be recreated a full wave EM simulator capable of providing detailed and accurate fields for electrically large complex regions with curvilinear surfaces is needed. The simulation architecture will be used to simulate the original scene and the final illusion within a complex environment. It should also further provide a means to address the problem of specifying the surface characteristics of the final illusion object, and compute the corresponding total fields.

Although numerical EM simulation techniques exist in many forms including volumetric methods such as  finite difference time and frequency domain approaches \cite{taflove2000computational}, it is integral equation approaches that are the most suitable for this problem\cite{chew2009integral}. These approaches are based on a discretization of the surfaces present in the problem and the determination of surface currents (virtual or real) that produce the appropriate fields in the simulation domain. Particularly appropriate to scattering problems with curvilinear surfaces in homogeneous regions, they are ideally placed to solve the initial object scattering problem, provide an infrastructure for the synthesis of the metasurface illusion object and confirm the effectiveness of the final illusion.

\begin{figure*}
    \centering
    \begin{subfigure}[b]{\columnwidth}
       \begin{overpic}[grid=false, scale = 0.22]{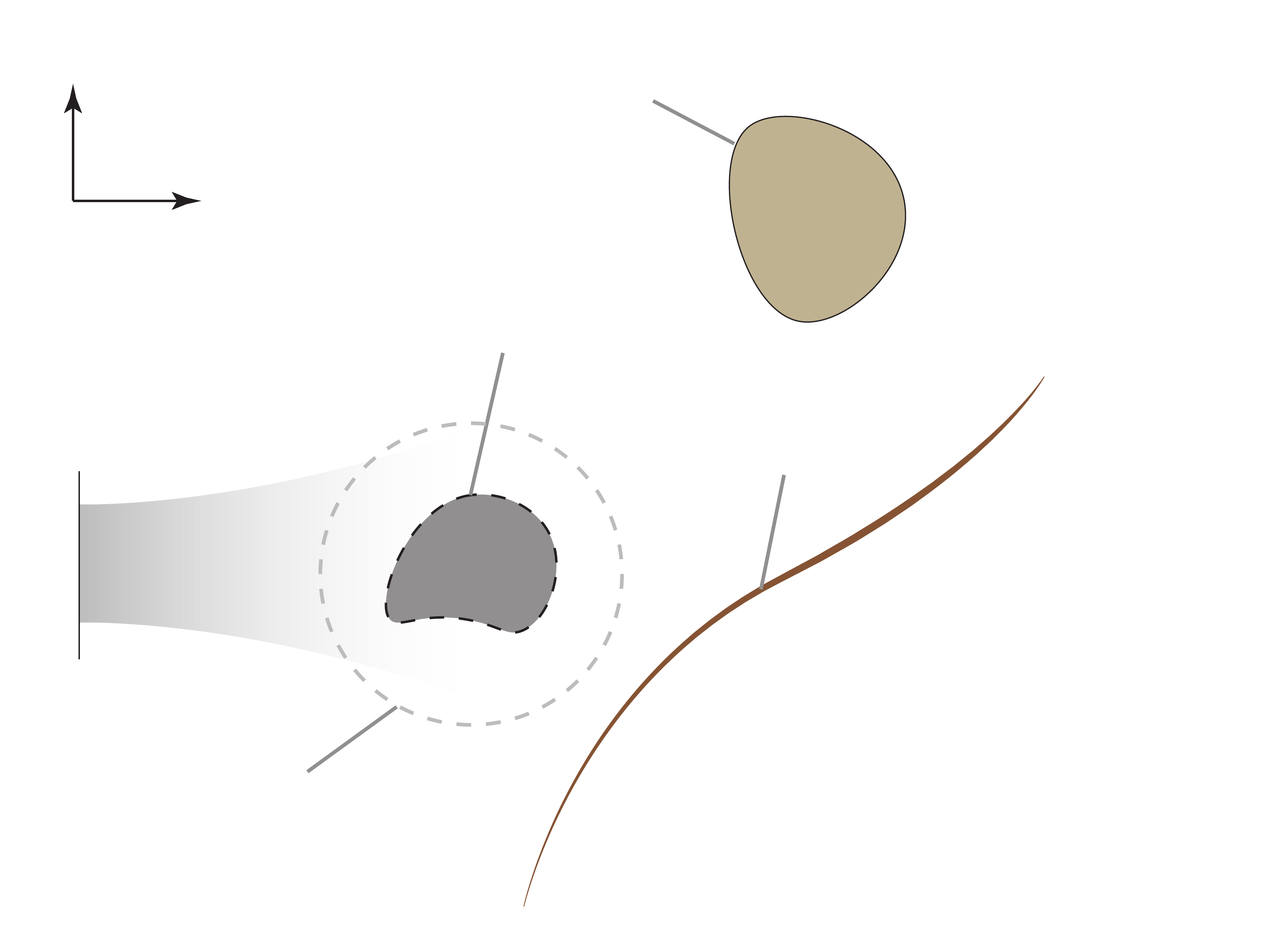}
        \put(5, 67){\scriptsize$x$}\put(17, 56){\scriptsize$y$}
        \put(-3, 36){\scriptsize{\shortstack{\textsc{Incident Field}\\ $\psi^\text{inc.}(\mathbf{r},\omega)$}}}
        \put(10, 2){\scriptsize{\shortstack{\textsc{Reference} \\ \textsc{Scattered Field}\\ $\psi_s^\text{ref}(\mathbf{r},\omega)$}}}
        \put(32, 47){\scriptsize{\color{ceruleanblue}\shortstack{\textsc{PEC Object (O)} \\ $ \Tv_{O} \Sv_O = \0$}}}
        \put(23, 65){\scriptsize{\color{ceruleanblue}\shortstack{\textsc{Dielectric Inclusion (D)}\\ $ \Dv_{D} \Sv_D = \0$}}}
        \put(47, 7){\scriptsize{\shortstack{\textsc{Opaque/Transmissive} \\ \textsc{Background Medium (B)}}}}
        \put(50,37){\scriptsize{\color{ceruleanblue}\shortstack{GSTCs \\ $\Dv^\times_{B} \Sv_B = \Gv_{B} \Sv_B$}}}

       \end{overpic}

 %\put(0,62){\scriptsize{\shortstack{\textsc{Total Fields}\\
       % $\psi^\text{inc.} + \psi_s^\text{ref} + \psi^{\text{ill}-}$}}}
%\put(){\shortstack{\textsc{External Illumination}\\ $\psi^{\text{ill}-}(\mathbf{r},\omega)$}}
%\psfrag{H}[c][c][0.7]{\shortstack{\textsc{Internal} \\ \textsc{Illumination}\\ $\psi_i^{\text{ill}+}(\mathbf{r},\omega)$}}
%\psfrag{B}[c][c][0.8]{\color{amber}\shortstack{\textsc{Metasurface}\\\textsc{Hologram/Skin (I)} \\ $ \Xv_I = \Qv^{-1} \Dv_{I}\Sv_{I}$}}
%\psfrag{x}[c][c][0.7]{$x$}
%\psfrag{y}[c][c][0.7]{$y$}
%\psfrag{F}[c][c][0.7]{\color{ceruleanblue}\shortstack{GSTCs \\ $\Dv^\times_{B} \Sv_B = \Gv_{B} \Sv_B$}}
%\psfrag{K}[c][c][0.7]{\color{ceruleanblue}\shortstack{GSTCs \\ $\Dv^\times_{I} \Sv_I = \Gv_{I} \Sv_I$}}
        \caption{}
    \end{subfigure}\quad
    \begin{subfigure}[b]{\columnwidth}
        \begin{overpic}[grid=false, scale = 0.22]{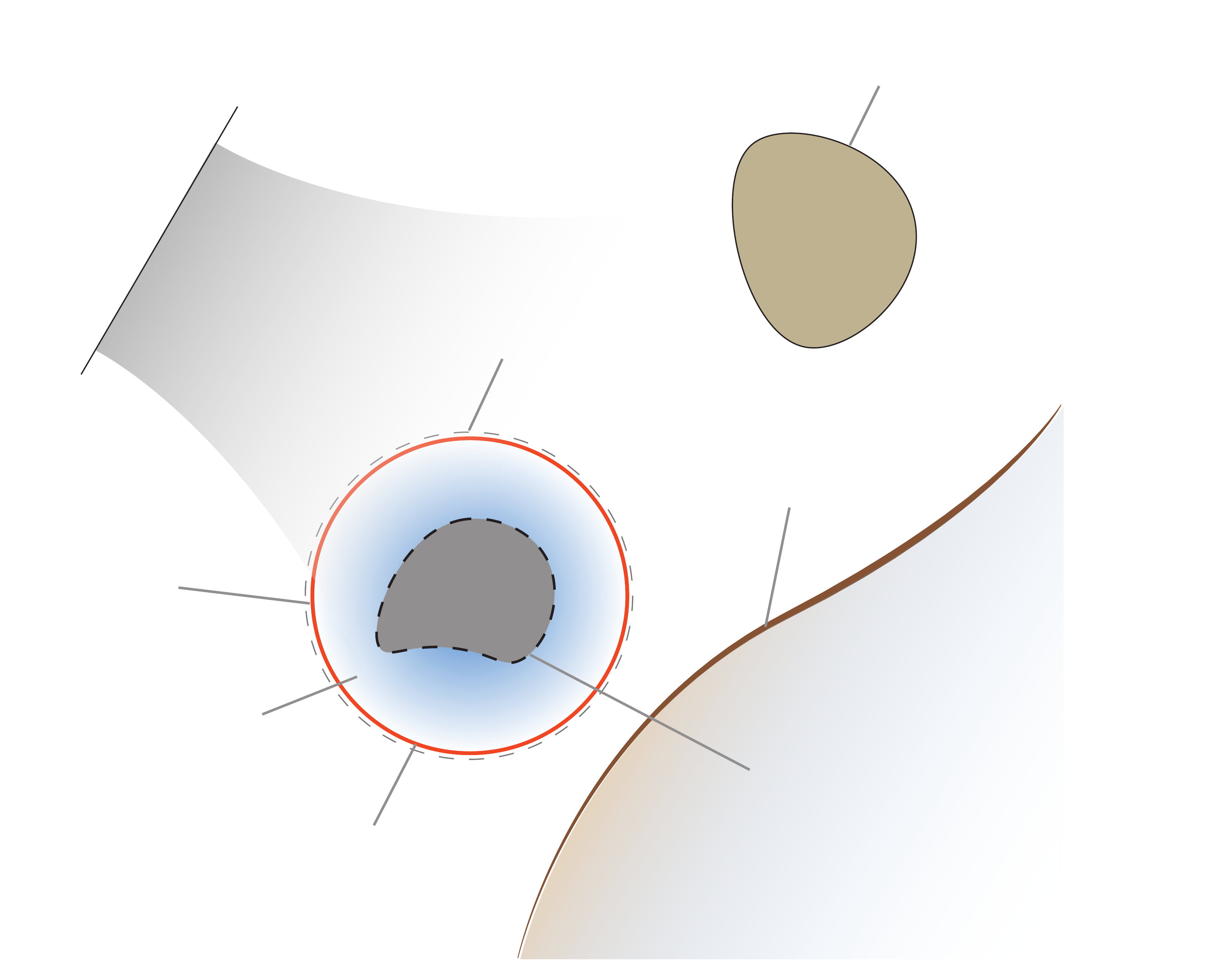}
            \put(11, 53){\scriptsize{\shortstack{\textsc{Total Fields}\\$\psi^\text{inc.}(\r_{m-}) + \psi_s^\text{ref}(\r_{m-}) + \psi^{\text{ill}-}(\r_{m-})$}}}
            \put(5,75){\scriptsize{\shortstack{\textsc{External Illumination}\\ $\psi^{\text{ill}-}(\mathbf{r},\omega)$}}}
            \put(-1,15){\scriptsize{\shortstack{\textsc{Internal} \\ \textsc{Illumination}\\ $\psi_i^{\text{ill}+}(\mathbf{r},\omega)$}}}
            \put(57,9.5){\scriptsize{\color{ceruleanblue}\shortstack{\textsc{PEC Object (O)} \\ $ \Tv_{O} \Sv_O = \0$}}}
            \put(58,75){\scriptsize{\color{ceruleanblue}\shortstack{\textsc{Dielectric Inclusion (D)}\\ $ \Dv_{D} \Sv_D = \0$}}}
            \put(12,2){\scriptsize{\color{amber}\shortstack{\textsc{Metasurface}\\\textsc{Hologram/Skin (I)} \\ $ \Xv_I = \Qv^{-1} \Dv_{I}\Sv_{I}$}}}
            \put(53,40){\scriptsize{\color{ceruleanblue}\shortstack{GSTCs \\ $\Dv^\times_{B} \Sv_B = \Gv_{B} \Sv_B$}}}
            \put(-8,31.5){\scriptsize{\color{ceruleanblue}\shortstack{GSTCs \\ $\Dv^\times_{I} \Sv_I = \Gv_{I} \Sv_I$}}}
            
        \end{overpic}

%\psfrag{D}[c][c][0.7]{\shortstack{\textsc{Incident Field}\\ $\psi^\text{inc.}(\mathbf{r},\omega)$}}
%\psfrag{J}[c][c][0.7]{\shortstack{\textsc{Reference} \\ \textsc{Scattered Field}\\ $\psi_s^\text{ref}(\mathbf{r},\omega)$}}
%\psfrag{E}[c][c][0.7]{\shortstack{\textsc{Opaque/Transmissive} \\ \textsc{Background Medium (B)}}}

\psfrag{x}[c][c][0.7]{$x$}
\psfrag{y}[c][c][0.7]{$y$}

        \caption{}
    \end{subfigure}
\caption{An illustration of a complex scene to be simulated consisting of various surfaces [Dielectric Inclusion (D), PEC Object (O) and Metasurface Background (B)] a) Reference simulation to determine the desired scattered fields. b) A closed metasurface with internal illumination either replacing (Hologram) the PEC object or enclosing it (camouflaging) to recreate the reference fields.}\label{Fig:SimSetup}
\end{figure*}

\subsection{Discretized IE Formulation}\label{Sec:III-A Discretized IE Formulation}

Consider a complex scene consisting of various type of objects and surfaces, for which the fields need to be solved, as shown in Fig.~\ref{Fig:SimSetup}(a). To formulate the problem as a solution to the integral formulation of Maxwell's equations the solution domain is broken into homogeneous regions with a constant index of refraction, separated by surfaces -- which can be real or fictitious. Each surface is characterized by a relationship between the tangential surface fields on both sides and an incident (forcing field) can be placed on appropriate surfaces. A complete solution will involve the incident fields at the surfaces ($\E^i/\H^i$)  and the scattered fields ($\E/\H$) from each surface determined by surface currents $\J$ and $\K$. Coupling between the regions is captured by the surface characterization. 

To make the problem numerically tractable the surfaces are discretized using elements of length $\delta \ell$. Defining for a particular surface ($S$) we have,
\begin{align*}
    \rbb_S = \bbmatrix \r_{S,1} & \dots & \r_{S,m} \ebmatrix, \quad \NN_S = \bbmatrix \Nh_{S,1} & \dots & \Nh_{S,m} \ebmatrix 
\end{align*}
where $\rbb_S$ denotes the center of a set of line segments and $\NN_S$ the element surface normals. On these surface elements we can define surface currents,
\begin{align*}
    \Jv_S = \bbmatrix \J_{S,1} & \dots & \J_{S,m} \ebmatrix, \quad \Kv_S = \bbmatrix \K_{S,1} & \dots & \K_{S,m} \ebmatrix 
\end{align*}
These surface current vectors can be grouped for convenience into the vector $\Cv$,
\begin{align*}
    \Cv_S = \bbmatrix \Jv_S & \Kv_S \ebmatrix^\top. 
\end{align*} 

\noindent Once this discretization has been defined, the modeling problem is composed of three related aspects; 1) Field propagation through the regions, 2) surface characterization and 3) coupling between the regions through the surfaces, which now will be treated separately as follows:

\noindent \textit{1. Field Propagation}

The EM fields radiated from each surface due to electric and magnetic surface currents can be generally expressed using a propagation matrix $\Pv_{S,p}$ as \cite{chew2009integral,
Method_Moments}:
\begin{align}\label{eq:EMProp}
    \bbmatrix \Ev(\rbb_p) \\\Hv(\rbb_p) \ebmatrix & = 
    \bbmatrix - j\omega \mu \Lv(\rbb_p, \rbb_S)  - \Rv(\rbb_p, \rbb_S)  \\ - j\omega \epsilon \Lv(\rbb_p, \rbb_S)  + \Rv(\rbb_p, \rbb_S)  \ebmatrix 
    \bbmatrix  \Jv_S \\ \Kv_S\ebmatrix \notag  \\
    \Fv_{p,S} & = \Pv_{S,p} \Cv_S
\end{align}
where $\rbb_p = \bbmatrix \r_{p,1} & \dots & \r_{p,n} \ebmatrix$ is a set of points at which the fields are desired, and $\rbb_S$ are the surfaces where the source currents are defined. The discretized field propagation matrices $\Lv$ and $\Rv$ are given by,
\begin{align*}
\Lv(\rbb_p, \rbb_S) &= \bbmatrix 
\L_{1,1} & \L_{1,2} & \dots &\L_{1,m} \\
\L_{2,1} & \L_{2,2} & \dots &\L_{2,m} \\
\vdots & \vdots & \ddots & \vdots\\
\L_{n,1} & \L_{n,2} & \dots &\L_{n,m} \\
\ebmatrix\\
\Rv(\rbb_p, \rbb_S) &= \bbmatrix 
\R_{1,1} & \R_{1,2} & \dots &\R_{1,m} \\
\R_{2,1} & \R_{2,2} & \dots &\R_{2,m} \\
\vdots & \vdots & \ddots & \vdots\\
\R_{n,1} & \R_{n,2} & \dots &\R_{n,m} \\
\ebmatrix
\end{align*}
with 
\begin{align*}
    \L_{i,j} &= \L(\rb_{p,i},\rb_{S,j})   \\
    &= \int_{\delta\ell_j}[1+\frac{1}{k^2}\nabla_{\rb_p}\nabla_{\rb_p}\cdotp] [G(\r_{p,i},\r_{S,j})] \,d\r_{S,j}\\
    \R_{i,j} &= \R(\rb_{p,i},\rb_{S,j})   
     = \int_{\delta\ell_j}\nabla_{\rb_p} \times [G(\r_{p,i},\r_{S,j})] \,d\r_{S,j}
\end{align*}
where  $G(\cdot)$ represents the Green's function. For a 2D case, the Green's function is given by the 2$^\text{nd}$ Hankel function, 
\begin{align*}
G(\r_S, \r_p)=    H_0^{(2)}(\r_S, \r_p) = J_0(\r_S, \r_p) - i Y_0(\r_S, \r_p),
\end{align*}
where $J_0(\cdot)$ and $Y_0(\cdot)$ are the Bessel functions of the 1$^\text{st}$ and 2$^\text{nd}$ kind representing outwardly propagating radial waves.

It will be useful to define propagation matrices for when fields are needed on the surface itself so that $\rbb_p = \rbb_S$. In such a case, we have for the two sides of the surface,
\begin{align*}
    \Fv_{S^+} = \Pv_{S,S^+} \Cv_S, \quad
    \Fv_{S^-}  = \Pv_{S,S^-} \Cv_S
\end{align*}
where $\{\cdot\}^+$ and $\{\cdot\}^-$ respectively, denote the right and left sides of an open surface or interior and exterior sides for a closed surface. Defining a surface field configuration $\Sv_S = \bbmatrix \Fv_{S^+} & \Fv_{S^-} \ebmatrix^T$ and a surface propagator $\Pv_S = \bbmatrix \Pv_{S^+} & \Pv_{S^-} \ebmatrix$, the above equation can be written in a compact matrix form as:
\begin{align} \label{eq:SurfPropBoth}
    \Sv_S = \Pv_S \Cv_S.
\end{align}

\noindent \textit{2. Surface characterization}

For each surface (open or closed) present in the domain, it is needed to formulate the surface field relationships that relate the total tangential fields on the two sides of the surface. Although a number of surface formulations can be accommodated within the framework, we will limit ourselves to \textit{dielectric boundaries, perfect electrical conductors (PEC) and metasurfaces described by the GTSCs}. Due to the flexibility of the GSTCs, this set of surfaces can describe a very wide range of possibilities in practice.  

The simplest surface is the PEC where the tangential electric field on both sides is equal to zero. This can be described by,
\begin{align}
\left[ \begin{array}{cccc}
     \NN_{T} & \0 & \0  & \0\\
     \0 &\0 &\NN_{T} & \0
\end{array}\right] \left[ \begin{array}{c}\Ev_{S^+}\\\Hv_{S^+}\\\Ev_{S^-}\\\Hv_{S^-}\end{array}\right]
 &= \left[ \begin{array}{cccc} \0 \\ \0 \end{array}\right]\notag\\
 \Tv_{S} \Sv_S &= \0
\end{align}
and the matrix operator $\NN_T$ performs the operation of extracting the two tangential fields at the surface (one in the $xy$ plane and the other with respect to $z$) obtaining $\E_T$ from $\E$ for every surface element.

For a dielectric surface the tangential fields on both sides are equal and we have,
\begin{align}
\left[ \begin{array}{cccc}
     \NN_{T} & \0 & -\NN_{T}  & \0\\
     \0 &\NN_{T} &\0 & -\NN_{T}
\end{array}\right] \left[ \begin{array}{c}\Ev_{S^+}\\\Hv_{S^+}\\\Ev_{S^-}\\\Hv_{S^-}\end{array}\right]
 &= \left[ \begin{array}{cccc} \0\\\0
\end{array}\right]\notag \\
 \Dv_{S} \Sv_S &= \0
\end{align}
where $\Dv_{S}$ takes the difference of the tangential fields for each surface element.

Finally, the most general surface description we use is for the metasurface.  Putting \eqref{Eq:GSTC} into a discrete form we obtain,
\begin{align}
&\left[ \begin{array}{cccc}
     \NN_{T\times} & \0 & -\NN_{T\times}  & \0\\
     \0 &\NN_{T\times} &\0 & -\NN_{T\times}
\end{array}\right] \left[ \begin{array}{c}\Ev_{S^+}\\\Hv_{S^+}\\\Ev_{S^-}\\\Hv_{S^-}\end{array}\right]
 = \notag\\
& \left[ \begin{array}{cccc}
     \gamma_\text{me}\NN_{T}  & \gamma_\text{mm}\NN_{T} & \gamma_\text{me}\NN_{T} & \gamma_\text{mm}\NN_{T}\\
     \gamma_\text{ee}\NN_{T}  & \gamma_\text{em}\NN_{T} & \gamma_\text{ee}\NN_{T} & \gamma_\text{em}\NN_{T}\\
\end{array}\right]
\left[ \begin{array}{c}\Ev_{S^+}\\\Hv_{S^+}\\\Ev_{S^+}\\\Hv_{S^+}\end{array}\right] \notag\\
&\Dv^\times_{S} \Sv_S = \Gv_{S} \Sv_S \label{Eq:gstcd}
\end{align}
\noindent where the surface susceptibility terms are expressed using auxiliary variables as,
\begin{align*}
    \gamma_\text{ee} = \frac{j\chi_\text{ee}\omega\epsilon}{2},~\gamma_\text{me/em} = \mp \frac{j\chi_\text{me/em}\omega\sqrt{\mu\epsilon}}{2},~\gamma_\text{mm} = -\frac{j\chi_\text{mm}\omega\mu}{2}.
\end{align*}
The operator $\NN_{T\times}$ extracts the total tangent field and then rotates these two fields to implement the $\Nh \times\{\cdot\}_T$ operation on every element and $\Dv^\times_{S}$ takes the difference of the rotated tangential fields.

\noindent \textit{3. System Level Simulation (Surface Couplings)}

The illustration example of Fig.~\ref{Fig:SimSetup}(a) consists of three types of objects in the incident field region: a dielectric inclusion (D), a closed PEC object (O) and a background metasurface (B)\footnote{Note that the background is modeled using GSTCs describing an arbitrarily complex surface, as a way to introduce sophisticated texture effects to be used in later examples.}. We therefore have three current vectors $C_\text{D}$, $C_\text{O}$ and $C_\text{B}$ which we can group into a single system level current vector,
\begin{align*}
        \Cv = \bbmatrix \Cv_\text{D} &\Cv_{O} &\Cv_\text{B} \ebmatrix^T
\end{align*}

\noindent For each surface we have propagators to other surfaces --  such as $\Pv_{\text{B,D}^-}$ which would determine the fields on the exterior side of the dielectric inclusion surface (D$^-$) due to the background surface (B). Using these propagators, we can formulate expressions for the fields at a surface as sums of propagated fields. For the dielectric inclusion, for instance, the internal fields (D$^+$) are due
to currents on the dielectric surface, while the external fields (D$^-$) are due to the currents on the dielectric itself, the PEC object (O) and the background surface (B), so that:
\begin{align*}
    \bbmatrix \Fv_{D^+} \\ \Fv_{D^-}\ebmatrix 
    &= \bbmatrix
    \Pv_{D,D^+} \Cv_D \\
    \Pv_{D,D^-} \Cv_D  + \Pv_{O,D^-} \Cv_O + \Pv_{B,D^-} \Cv_B  \\
    \ebmatrix
\end{align*}
We can then define a system level propagator for each surface which relates the surface fields to all the currents on various objects, such as:
\begin{align*}
    \Sv_D &=  \bbmatrix
    \Pv_{D,D^+} &\0 &\0\\
    \Pv_{D,D^-} & \Pv_{O,D^-} & \Pv_{B,D^-}   \\
    \ebmatrix
    \bbmatrix  \Cv_D\\\Cv_E\\\Cv_B  \ebmatrix = \Pv_D \Cv.
\end{align*}
In a similar manner we can define the system level propagators, $\Pv_O$ and $\Pv_B$, for the PEC object and the background, respectively. For each surface we must also define the surface relationships for the fields i.e,
\begin{align*}
\begin{array}{ll}
    \Tv_O \Sv_O = \0 & \text{(PEC Object)}\\
    \Dv_D \Sv_D = \0  & \text{(Dielectric Inclusion)}\\
    \Dv^\times_B \Sv_B = \Gv_B \Sv_B & \text{(Metasurface Background)}
\end{array}
\end{align*}
\noindent Finally, to complete the physical specification of the region equations, we must ensure that the current only flows on the surface by taking the dot product of the surface normal with the currents on the surface:
\begin{align*}
% \begin{array}{lll}
    \{ \Nv_D \Cv_O = \0, \; \Nv_D \Cv_D = \0, \;  \Nv_D \Cv_B = \0 \}  \; \Rightarrow \;\Nv_D \Cv = \0,
% \end{array}
\end{align*}
where the operator $\Nv_D$ performs a dot product for currents on every element. 

All of these equations can now be assembled into a matrix equation where the unknowns are the surface currents and the fields:
\begin{align}\label{Eq:SystemMatrix}
    \bbmatrix
    \Pv_D & \0 & \I & \0 & \0 \\
    \0 & \0 & \Dv_D & \0 & \0 \\
    \Pv_O & \0 & \0 & \I & \0 \\
    \0 & \0 & \Tv_O & \0 & \0 \\
    \Pv_B & \0 & \0 & \0 & \I \\
    \0 & \0 & \0 & \Dv^\times_B - \Gv^\times_B \\
    \Nv_D & \0 & \0 & \0 & \0 \\
    \ebmatrix
    \bbmatrix \Cv \\ \Sv_D \\ \Sv_O \\ \Sv_B \ebmatrix
     = 
    \bbmatrix \0 \\ -\Dv_D\Sv^\text{inc.}_D \\ \Tv_O\Sv^\text{inc.}_B\\ -(\Dv_B-\Gv_B)\Sv^\text{inc.}_B \ebmatrix
\end{align}
\noindent where an applied incident field $\Sv^\text{inc.}$ is present on all surfaces. This system of equations can now be solved and then the fields anywhere in the region due to newly found currents are computed using various propagation matrices.

\subsection{Illusion Design}\label{Sec-III-B:Illusion Design}

Let us consider the objective where we wish to remove the PEC object of interest by a metasurface hologram as shown in Fig.~\ref{Fig:SimSetup}(b), which is designed in such a way that the metasurface creates identical reference fields, [incident $\psi^\text{inc.}(\r,\omega)$ + scattered $\psi_s^\text{ref.}(\r,\omega)$, in general], at the observer compared to that when the object was present, i.e. object illusion. The first step in the illusion design is to compute the desired fields (i.e. reference fields/scene) to be recreated by the closed metasurface at and beyond its location. The surface of the closed metasurface hologram (also referred to as the illusion surface $I$, here) can be discretized as:
\begin{align*}
    \rbb_I = \bbmatrix \r_{I,1} & \dots & \r_{I,m} \ebmatrix
\end{align*}
We can create a propagation matrix $\Pv_I$ that calculates the \textit{scattered fields} $\Fv_I$ present in the reference scene (without the metasurface hologram) at the illusion surface,
\begin{align}
    \Fv_I^\text{ref} = \Pv_{I}\Cv \label{Eq:ReferenceFields}
\end{align}
where $\Cv$ represents all the currents present in the original reference scene. Next step is to remove the PEC object and introduce the metasurface hologram producing the illusion.

To synthesize the illusion surface we need to determine the total fields present on both sides of the surface. Let us consider the general case, where both internal ($\Fv_{I}^{\text{ill}+}$) and external ($\Fv_{I}^{\text{ill}-}$) illumination fields are present and the illumination field is different from the incident field $\Fv_{I}^\text{inc.}$. The fields in the external region consists of scattered fields we wish to recreate along with the original incident field to which we must add the external illumination, so that 
\begin{align}
    \Fv_I^- = \Fv_I^\text{ref} + \Fv_{I}^\text{inc.} + \Fv_{I}^{\text{ill}-}\label{Eq:ExF_ill_ne_Inc}
\end{align}
If the external illumination is chosen to be the same as the incident fields in the reference simulation, then the total external fields simply becomes $ \Fv_I^- = \Fv_I^\text{ref} + \Fv_{I}^\text{inc.}$.
\begin{align}
    \Fv_I^- = \Fv_I^\text{ref} + \Fv_{I}^\text{inc.}\label{Eq:ExF_ill_e_Inc}
\end{align}

With respect to the scattered fields interior to the surface, on the other hand, have some design flexibility. While the exact choice will not impact the reconstruction of the desired scattered fields, it will have a significant effect on the required surface susceptibilities of the metasurface hologram. For simplicity, let us set all the internal scattered fields from the metasurface to zero, so that the total internal field is simply given by the internal illumination as  
\begin{align*}
    \Fv_I^- = \Fv_{I}^{\text{ill}+}
\end{align*}
We can then form the surface field vector as
\begin{align*}
    \Sv_I = \bbmatrix \Fv_I^\text{ref} + \Fv_{I}^\text{inc.} + \Fv_{I}^{\text{ill}-}\\ \Fv_{I}^{\text{ill}+} \ebmatrix.
\end{align*}

The internal illumination may be seen as an extra degree of control available to the system designer. In the case of creating an electromagnetic illusion, the metasurface hologram is hollow, and the internal source may be placed anywhere inside the structure. It could be as simple as a radially propagating wave from the center, for instance, which can be modeled as a Henkel function of the $2^\text{nd}$ kind producing a uniform field amplitude at the interior surface of the metasurface shield, i.e. 

\begin{align}
E_z(\r_S, \r_m^+) =    H_0^{(2)}(\r_S, \r_m^+),
\end{align}

\noindent where $\r_S$ is the source location of the internal illumination, typically at the center of the circular metasurface hologram assumed here, and $\r_m^+$ are the locations at the metasurface internal to it where the fields are computed. In the case of an electromagnetic camouflage, the metasurface will enclose the target object, and the internal source must now be engineered to produce the same uniform internal illumination at the interior surface of the metasurface, but in the presence of the object this time. We will hereafter assume that such an internal illumination is available, for either of the application cases of illusion or camouflage to simplify the subsequent synthesis procedure.

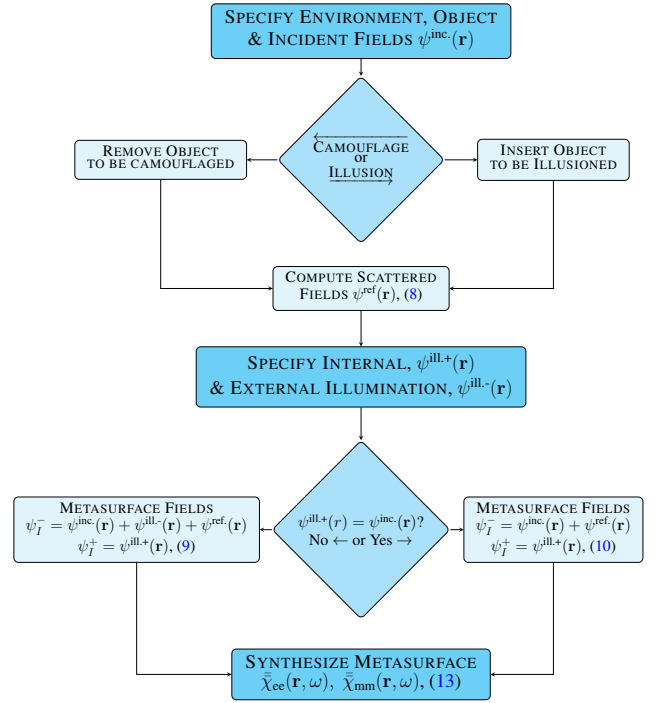
\begin{figure}[htbp]
\begin{center}
\resizebox{\columnwidth}{!}{ 
\tikzstyle{startstop} = [rectangle, rounded corners, minimum width=3cm, minimum height=1cm,text centered, draw=black, fill=cyan!50]
\tikzstyle{io} = [trapezium, trapezium left angle=70, trapezium right angle=110, minimum width=3cm, minimum height=1cm, text centered, draw=black, fill=blue!30]
\tikzstyle{process} = [rectangle, rounded corners, minimum width=3cm, minimum height=1cm, text centered, draw=black, fill=cyan!10]
\tikzstyle{decision} = [diamond, rounded corners, minimum width=3cm, minimum height=1cm, text centered, draw=black, fill=cyan!30]
\tikzstyle{arrow} = [thick, round cap->,>=stealth]
%\clearpage
\centering
%\footnotesize
\Large
\begin{tikzpicture}[node distance=2cm] 
\LARGE\node (start) [startstop] {\begin{tabular}{c}\shortstack{\textsc{Specify Environment, Object}\\ {\& \textsc{Incident Fields} $\psi^\text{inc.}(\r)$}}\end{tabular}};\Large
%\large
 \node (dec1) [decision, below of=start, yshift=-2.0cm] {\shortstack{$\overleftarrow{\textsc{Camouflage}}$ \\ or\\ $\underrightarrow{\textsc{Illusion}}$}};
\node (pro2a) [process, right of=dec1, xshift=4cm] {\begin{tabular}{c}\shortstack{\textsc{Insert Object} \\ \textsc{to be Illusioned}}\end{tabular}};
\node (pro2b) [process, left of=dec1, xshift=-4.25cm] {\begin{tabular}{c}\shortstack{\textsc{Remove Object} \\ \textsc{to be camouflaged}}\end{tabular}};

\node (pro3a) [process, below of=dec1, yshift=-2cm] {\begin{tabular}{c}\shortstack{\textsc{Compute Scattered} \\ \textsc{Fields} $\psi^\text{ref}(\r)$,~\eqref{Eq:ReferenceFields}}\end{tabular}};

\LARGE\node (pro4a) [startstop, below of=pro3a, yshift=-0.75cm] {\begin{tabular}{c}\shortstack{\textsc{Specify Internal}, $\psi^\text{ill.+}(\r)$\\ \textsc{\& External Illumination}, $\psi^\text{ill.-}(\r)$}\end{tabular}};\Large
\node (dec2) [decision, below of=pro3a, yshift=-5.5cm] {\shortstack{$\psi^\text{ill.+}(r)=\psi^\text{inc.}(\r)$? \\ No $\leftarrow$ or Yes $\rightarrow$}};
\node (pro5a) [process, right of=dec2, xshift=4cm] {\begin{tabular}{c} \shortstack{\textsc{Metasurface Fields}\\ $\psi_I^- = \psi^\text{inc.}(\r) + \psi^\text{ref.}(\r)$\\ $\psi_I^+ = \psi^\text{ill.+}(\r)$,~\eqref{Eq:ExF_ill_e_Inc}}\end{tabular}};
\node (pro5b) [process, left of=dec2, xshift=-5cm] {\begin{tabular}{c}\shortstack{\textsc{Metasurface Fields}\\ $\psi_I^- = \psi^\text{inc.}(\r) + \psi^\text{ill.-}(\r) + \psi^\text{ref.}(\r)$\\ $\psi_I^+ = \psi^\text{ill.+}(\r)$,~\eqref{Eq:ExF_ill_ne_Inc}}\end{tabular}};
\LARGE\node (pro6a) [startstop, below of=dec2, yshift=-2.5cm] {\begin{tabular}{c} \shortstack{\textsc{Synthesize Metasurface}\\ {$\bar{\bar{\chi}}_\text{ee}(\r, \omega),~\bar{\bar{\chi}}_\text{mm}(\r, \omega)$,~\eqref{Eq:Ms_Synth}}}\end{tabular}};\Large
\draw [arrow] (start) -- (dec1);
\draw [arrow] (dec1) -- (pro2a);
\draw [arrow] (dec1) -- (pro2b);
\draw [arrow] (pro2a) |- (pro3a);
\draw [arrow] (pro2b) |- (pro3a);
\draw [arrow] (pro3a) -- (pro4a);
\draw [arrow] (pro4a) -- (dec2);
\draw [arrow] (dec2) -- (pro5a);
\draw [arrow] (dec2) -- (pro5b);
\draw [arrow] (pro5a) |- (pro6a);
\draw [arrow] (pro5b) |- (pro6a);
\end{tikzpicture}}
\caption{Flowchart illustrating the design process of synthesizing the metasurface as a hologram or a skin for electromagnetic illusion or camouflage in a given scene.}\label{Fig:Flow}
\end{center}
\end{figure}

The metasurface surface susceptibility synthesis next rests on solving the GSTCs matrix equation of Eq.~\ref{Eq:gstcd} for the illusion surface as described by the matrices $\Dv^\times_{I}$ and $\Gv_{I}$ for the prescribed field $\Sv_I$. To extract the unknown surface susceptibilities we rearrange $\Gv_{I}$ as,
\begin{align*}
&\Gv_{I} = \left[ \begin{array}{cccc}
     \chi_\text{me}\Av_\text{me} & \chi_\text{mm}\Av_\text{mm}  & \chi_\text{me}\Av_\text{me} & \chi_\text{mm}\Av_\text{mm}\\
     \chi_\text{ee}\Av_\text{ee} & \chi_\text{em}\Av_\text{em} & \chi_\text{ee}\Av_\text{ee} & \chi_\text{em}\Av_\text{em}\\
\end{array}\right]
\end{align*}
with
\begin{align*}
  &  \Av_\text{ee} =  \frac{j\Nv_{T}\omega\epsilon}{2},~ \Av_\text{em/me} = \pm \frac{j\Nv_{T}\omega\sqrt{\mu\epsilon}}{2},~\Av_\text{mm} = -\frac{j\Nv_{T}\omega\mu}{2} 
\end{align*}
Considering that the metasurface susceptibilities are discretized over the surface as $\chi(\r_m)$ and thus become vectors of localized susceptibilities ($\Xv$), we can express the right hand side of Eq.~\ref{Eq:gstcd}, as
\begin{align*}
    \Gv_{I}\Sv_{I} 
    % &= \bbmatrix
    %   \chi_\text{me}\Av_{me} & \chi_\text{mm}\Av_{mm}  & \chi_\text{me}\Av_{me} & \chi_\text{mm}\Av_{mm}\\
    %  \chi_\text{ee}\Av_{ee} & \chi_\text{em}\Av_{em} & \chi_\text{ee}\Av_{ee} & \chi_\text{em}\Av_{em}\\
    % \ebmatrix 
    % \bbmatrix \Ev_1\\\Hv_1\\\Ev_2\\\Hv_2\ebmatrix\\
    % &= \bbmatrix
    %   \chi_\text{me}\Av_{me}\Ev_1 + \chi_\text{mm}\Av_{mm}\Hv_1  + \chi_\text{me}\Av_{me}\Ev_2 + \chi_\text{mm}\Av_{mm}\Hv_2\\
    %  \chi_\text{ee}\Av_{ee}\Ev_1 + \chi_\text{em}\Av_{em}\Hv_1 + \chi_\text{ee}\Av_{ee}\Ev_2 + \chi_\text{em}\Av_{em}\Hv_2
    % \ebmatrix \\
    &= \left[ \begin{array}{l}
      \Xv_\text{me}\circ\left(\Av_\text{me}(\Ev_1 + \Ev_2)\right) \; + \\ \quad \quad \Xv_\text{mm}\circ\left(\Av_\text{mm}(\Hv_1 + \Hv_2)\right)\\
     \Xv_\text{ee}\circ\left(\Av_\text{ee}(\Ev_1 + \Ev_2)\right)  \; + \\ \quad \quad \Xv_\text{em}\circ\left(\Av_\text{em}(\Hv_1 +  \Hv_2)\right)
    \end{array}\right],
\end{align*}
\noindent where $\circ$ is the point-wise Hagamard product. Each of the terms,
\begin{align*}
    \Av_\text{me}(\Ev_1 + \Ev_2) 
    % = \bbmatrix \E_{Gme,1}\\\vdots\\ \E_{Gme,N}\\\hline \E_{Gme,N+1} \\ \vdots \\ \E_{Gme,2N} 
    % \ebmatrix 
    = \bbmatrix \Bv_{\text{me},xy}\\\Bv_{\text{me},z} \ebmatrix = \Bv_\text{me},
\end{align*}
\noindent is a column vector of one component of tangent fields ($xy$ and $z$). If we wish to create a distributed $\chi$ vector we can form (for example) products such as,
\begin{align*}
    \Gv_\text{me} \Xv_\text{me} &= \Xv_\text{me} \circ \Av_\text{me}(\Ev_1 + \Ev_2)   
\end{align*}
\noindent where we define a diagonal matrix,
\begin{align*}
    \Gv_\text{me} = \bbmatrix 
    B_\text{me,1} & 0 & \dots & 0\\ 
    0  &B_\text{me,2} & \dots & 0 \\
    0  & 0 & \ddots & 0 \\
    0  & 0 & 0 & B_\text{me,2N}
    \ebmatrix
\end{align*}
\noindent This form allows a very convenient expression of RHS of Eq.~\ref{Eq:gstcd}(a), where the susceptibility matrix term is explicitly extracted as
\begin{align}
    \Gv_{I}\Sv_{I} 
    % &= \bbmatrix
    %   \chi_\text{me}\Av_{me}(\Ev_1 + \Ev_2) + \chi_\text{mm}\Av_{mm}(\Hv_1 + \Hv_2)\\
    %  \chi_\text{ee}\Av_{ee}(\Ev_1 + \Ev_2)  + \chi_\text{em}\Av_{em}(\Hv_1 +  \Hv_2)
    % \ebmatrix \\
    % &\bbmatrix
    %   \Bv_{me} \Xv_{em} +\Bv_{mm} \Xv_{mm}\\
    %   \Hv_{Gee} \Xv_{ee} +\Hv_{Gme} \Xv_{me}
    % \ebmatrix 
    &= 
    \bbmatrix
       \Gv_{me} & \Gv_{mm} &\0 & \0\\
      \0 & \0 &  \Gv_{ee} & \Gv_{me} 
    \ebmatrix
    \bbmatrix\Xv_{em}\\\Xv_{mm}\\\Xv_{ee} \\ \Xv_{me}\ebmatrix_I \nonumber\\
    &= \Qv \Xv_I.
\end{align}
 Finally using Eq.~\ref{Eq:gstcd}, we now have the explicit relationship for the spatially varying surface susceptibility matrix as
\begin{align}
    \Xv_I = \Qv^{-1} \Dv_{I}\Sv_{I},\label{Eq:Ms_Synth}
\end{align}
which can be used directly for metasurface synthesis for a given $\Sv_{I}$. This finally completes the synthesis of the metasurface shield. This entire design flow is illustrated in the flow chart of Fig.~\ref{Fig:Flow}. 

It should be noted that there is no unique surface susceptibility distribution that can generate the specified illusion. There are in total 36 complex unknowns in the susceptibility tensors, and the number of possible solutions generating the same illusion is virtually unlimited. Certain constraints may be applied such as reciprocity and lossless-ness, which may limit the number of combinations and may provide susceptibility distributions which may be more convenient to practically implement, i.e.
        %
        % \begin{align}
        % \text{Reciprocity:}&~\overline{\overline{\chi}}_\text{ee}^\text{T} =  \overline{\overline{\chi}}_\text{ee},\quad \overline{\overline{\chi}}_\text{mm}^\text{T} =  \overline{\overline{\chi}}_\text{mm},\quad \overline{\overline{\chi}}_\text{me}^\text{T} =  -\overline{\overline{\chi}}_\text{em}\\
        % \text{Losslessness:}&~\overline{\overline{\chi}}_\text{ee}^\text{T} =  \overline{\overline{\chi}}_\text{ee}^\ast,\quad \overline{\overline{\chi}}_\text{mm}^\text{T} =  \overline{\overline{\chi}}_\text{mm}^\ast,\quad \overline{\overline{\chi}}_\text{me}^\text{T} =  \overline{\overline{\chi}}_\text{em}^\ast.
        % \end{align}
        %
        \begin{align*}
        \begin{array}{lccc}
        \text{Reciprocity:} & ~\overline{\overline{\chi}}_\text{ee}^\text{T} =  \overline{\overline{\chi}}_\text{ee}, & \overline{\overline{\chi}}_\text{mm}^\text{T} =  \overline{\overline{\chi}}_\text{mm}, & \overline{\overline{\chi}}_\text{me}^\text{T} =  -\overline{\overline{\chi}}_\text{em}\\
        \text{Losslessness:} & ~\overline{\overline{\chi}}_\text{ee}^\text{T} =  \overline{\overline{\chi}}_\text{ee}^\ast, & \overline{\overline{\chi}}_\text{mm}^\text{T} =  \overline{\overline{\chi}}_\text{mm}^\ast,& \overline{\overline{\chi}}_\text{me}^\text{T} =  \overline{\overline{\chi}}_\text{em}^\ast.
        \end{array}
        \end{align*}
        In many practical cases, a fully general bi-anisotropic metasurface may not be desired (i.e. $\overline{\overline{\chi}}_\text{me} = \overline{\overline{\chi}}_\text{em}=0$), which will reduce the unknowns to 18. If further simplicity is desired, tensors maybe reduced to scalars ($\overline{\overline{\chi}}_\text{ee} = \chi_\text{ee}$ and $\overline{\overline{\chi}}_\text{mm} = \chi_\text{ee}$), with only 2 unknowns, as will be assumed here, which is the minimum number of unknowns to have unique solution. If material bounds must be respected, such as synthesizing lossless metasurfaces, scalar surface susceptibilities are not be sufficient, and extra elements of the tensorial susceptibilities must instead be used, which can easily be incorporated inside the proposed method.
        
Given that we can now synthesize an illusion surface, we turn to a more precise definition of the \textit{Observer}, before providing numerical demonstration of the proposed synthesis method.
 
 \begin{figure*}[h]
\begin{center}
	\begin{overpic}[grid=false, scale = 0.27]{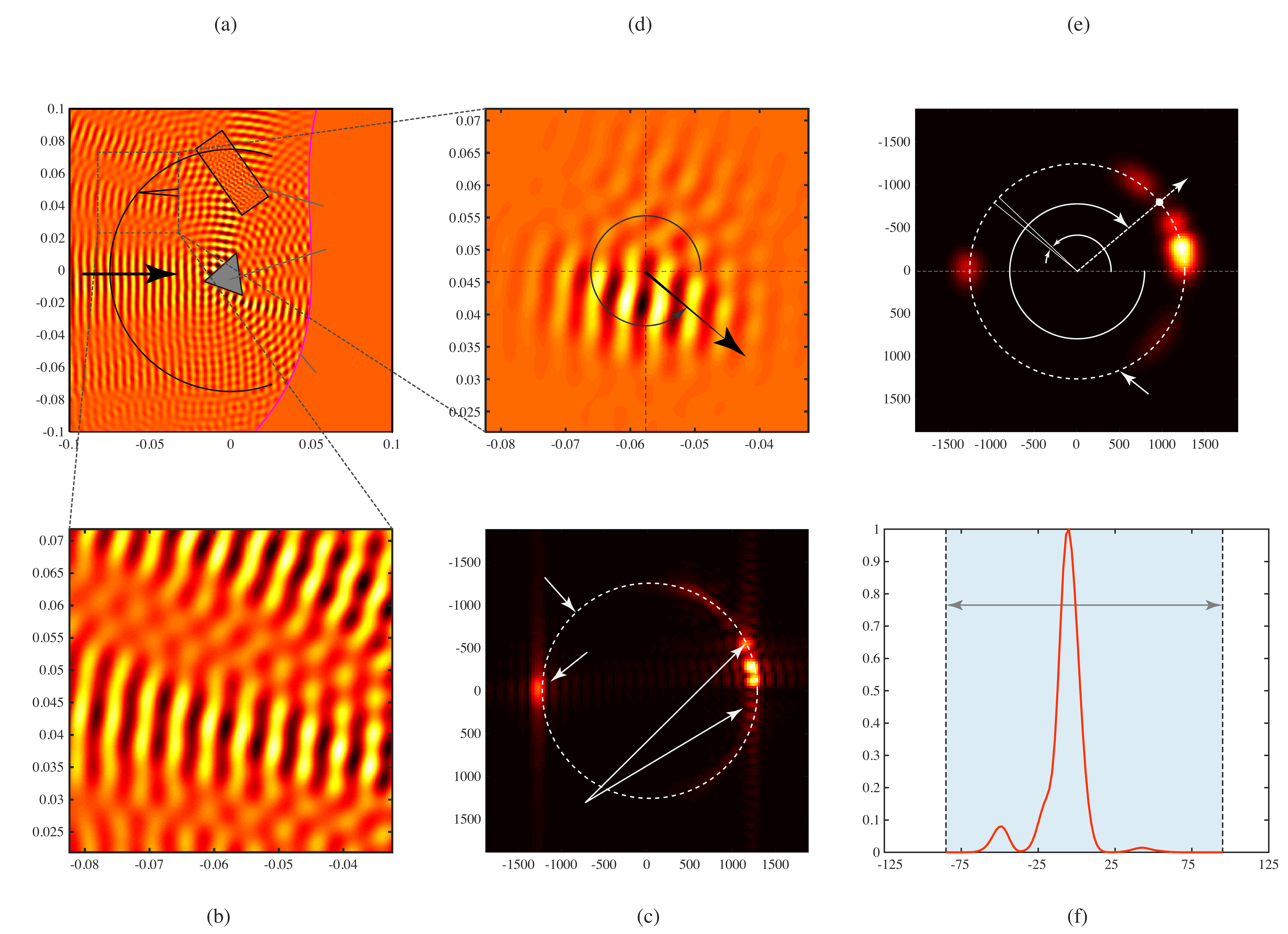}
	    \put(17,36.5){\scriptsize{$x$~(m)}}
	 \put(1,50){\rotatebox{90}{\scriptsize{$y$~(m)}}}
	 \put(10,66){\scriptsize{Total Fields Re$\{E_z^\text{tot.}(x,y)\}$}}
	 \put(25.5,55.5){\scriptsize{Dielectric}}
     \put(25.5,53){\scriptsize{PEC}}
     \put(22,42.5){\scriptsize{Background}}
     \put(6,49){\scriptsize{Incident Fields}}
	 
	    \put(49,36.5){\scriptsize{$x$~(m)}}
	\put(33,50){\rotatebox{90}{\scriptsize{$y$~(m)}}}
	 \put(39,65){\scriptsize{\shortstack{$E_z^\text{obs.} = $ Re$\{E_z^\text{tot.}(x,y)\}\times w(x,y)$\\ Gaussian Weighting}}}
	 \put(51.5,56.5){\scriptsize{$\theta$}}
     \put(55,43){\scriptsize{POV,~$\mathcal{P}(\theta)$}}
	
        \put(81,36){\scriptsize{$k_x$~(rad/m)}}
     \put(66,49){\rotatebox{90}{\scriptsize{$k_y$~(rad/m)}}}
      \put(75,66){\scriptsize{$\tilde{E}_z^\text{tot.}(k_x, k_y) = \mathcal{F}\{E_z^\text{obs.}(x,y)\}$}}
     \put(82.5,55){\scriptsize{\color{white}$\phi$}}
     \put(80,50){\scriptsize{\color{white} $d\phi$}}
     \put(89,41){\scriptsize{\color{white} $\rho = k_0$}}
     \put(88,52.5){\scriptsize{\color{white}$\theta$}}
     
	    \put(17,3){\scriptsize{$x$~(m)}}
    \put(1,17.5){\rotatebox{90}{\scriptsize{$y$~(m)}}}
     \put(7,32.5){\scriptsize{\shortstack{$E_z^\text{obs.} = $ Re$\{E_z^\text{tot.}(x,y)\}\times w(x,y)$\\ Rect Weighting}}}
     
        \put(47,3){\scriptsize{$k_x$~(rad/m)}}
      \put(33,15.5){\rotatebox{90}{\scriptsize{$k_y$~(rad/m)}}}
    \put(40.5,32.5){\scriptsize{$\tilde{E}_z^\text{tot.}(k_x, k_y) = \mathcal{F}\{E_z^\text{obs.}(x,y)\}$}}
      \put(40,8){\scriptsize{\color{white}Scattered Fields}}
     \put(43,22.5){\scriptsize{\color{white}Incident Fields}}
     \put(39,28.5){\scriptsize{\color{white} $\rho = k_0$}}
      
      \put(77,3){\scriptsize{Angular Scan $\theta$ (deg)}}
     \put(65,17.5){\rotatebox{90}{\scriptsize{$\mathcal{P}(\theta)$}}}
     \put(84,20){\scriptsize{\shortstack{Scene\\ $\theta\in\{-90^\circ-90^\circ\}$\\ $k_x>0,~k_y>0$}}}
     
     %\psfrag{Q}[c][c][0.7]{\color{white} $\theta$}
      
	\end{overpic}
     \caption{The proposed Structured Field Observation (SFO) method to construct a scene from the perspective of an observer. a) An example scene to be rendered. b) Sampled fields, $\psi^\text{obs.}(\r)$ in the observation region around the observer, using \eqref{Eq:RectSample}. c) 2D-FFT of the sampled fields using \eqref{Eq:RectSampleFFT}. d) Conditioned fields using a Gaussian weighting function of \eqref{Eq:GaussWeightFctn} and e) the corresponding 2D-FFT. f) Rendered image of the scene as described by the observed power as function of angle using \eqref{Eq:PowerScan}. The simulation parameters are: Dielectric inclusions size: $0.05~\text{m} \times 0.02~\text{m}$ refraction index, $n = 2$, PEC triangle object: 0.025~m side. Incident field is a Gaussian beam of width 0.02~m and the observer is located at: $(-0.057,0.048)$~m. }
     \label{fig:ObsDef}
\end{center}     
 \end{figure*}
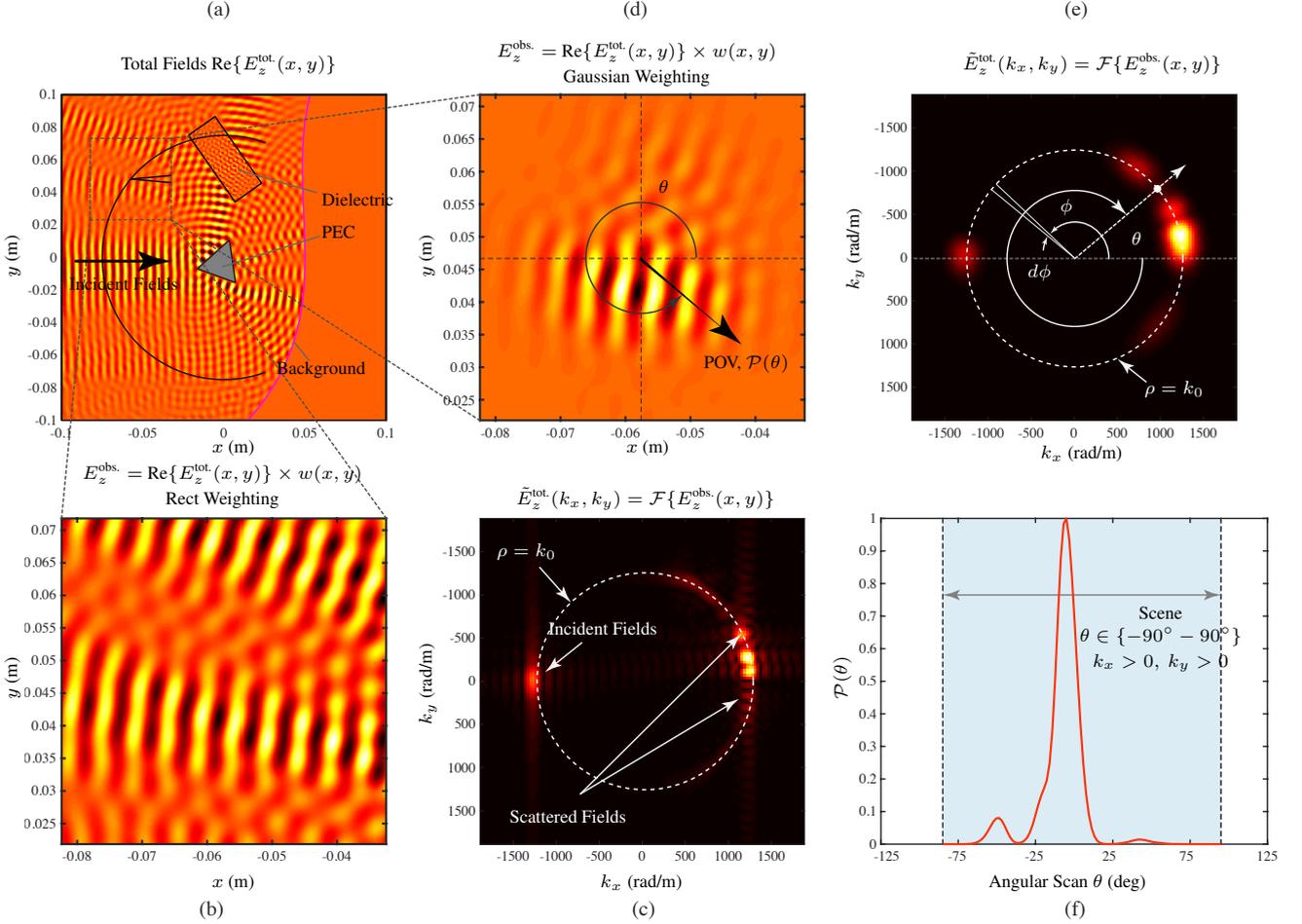

\section{Modeling an Observer}\label{Sec-IV:Modeling An Observer}

In order to evaluate the accuracy of an illusion or its effectiveness under various circumstances, such as environmental changes or illumination variation -- it is needed to compare the electromagnetic characteristics of the reference scene and the illusion scene. Two approaches are used in this paper: 1) a direct comparison of scattered or total fields at an observation position or a set of observation positions defined by a line or arc, and 2) a more sophisticated sampling of the field within an observation region allowing for the incorporation of a field of view and the directionality of the scattered and total field components. The second method can be used to, in essence, render an image of the scene.

\subsection{Direct Field Comparison}

A straight-forward method of evaluating the effectiveness of an illusion is to compare, at prescribed locations, the field distributions within the scene to the original baseline simulation. We will refer to this as Direct Field Observation (DFO). If wished, the complex vectorial field components can be compared or simply the magnitude of the various fields. Although this comparison is easy to accomplish there are a number of complications in interpretation. As discussed previously, if the observer is viewing a front-lit object from the front only the scattered field will be observed as the illuminating (or incident) field is propagating from behind the observer. On the other hand if viewing from the far side of a front lit object the observer will collect contributions from both the scattered field and the illuminating field. For an illusion created from a vertical metasurface as discussed in \cite{smy2020surface} these distinctions are easily handled by delineating illusions as front-lit or back-lit and subject to front or back illumination, however, for closed surface illusions, these distinctions are not clear. It is not obvious from inspection if total or scattered fields should be compared at a particular observation position. 

Another significant issue with this method is the lack of any ability to obtain information about the directional structure of the field in the region nearby the observation position. Without this information, the Observer cannot be characterized by a Point-of-View (POV) and a Field-of-View (FOV) -- the direction in which the observer is looking and the range of angles over which the observation is obtained. This limitation is a direct consequence of sampling the field with a delta function and only obtaining field information at a point. As a result of this limitation there is no ability to reconstruct the scene from the view point of the observer.

 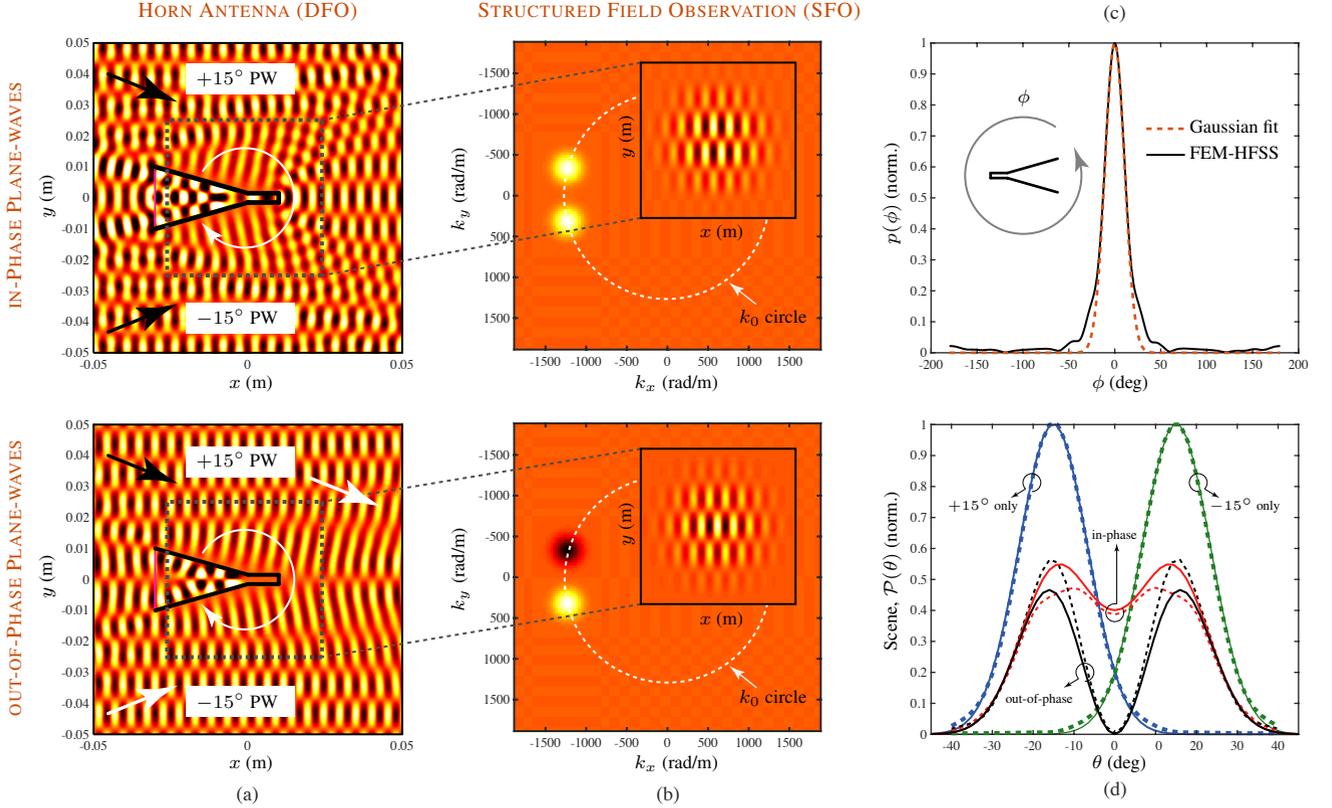
\begin{figure*}[!t]
     \centering
     \begin{overpic}[grid=false, scale = 0.25]{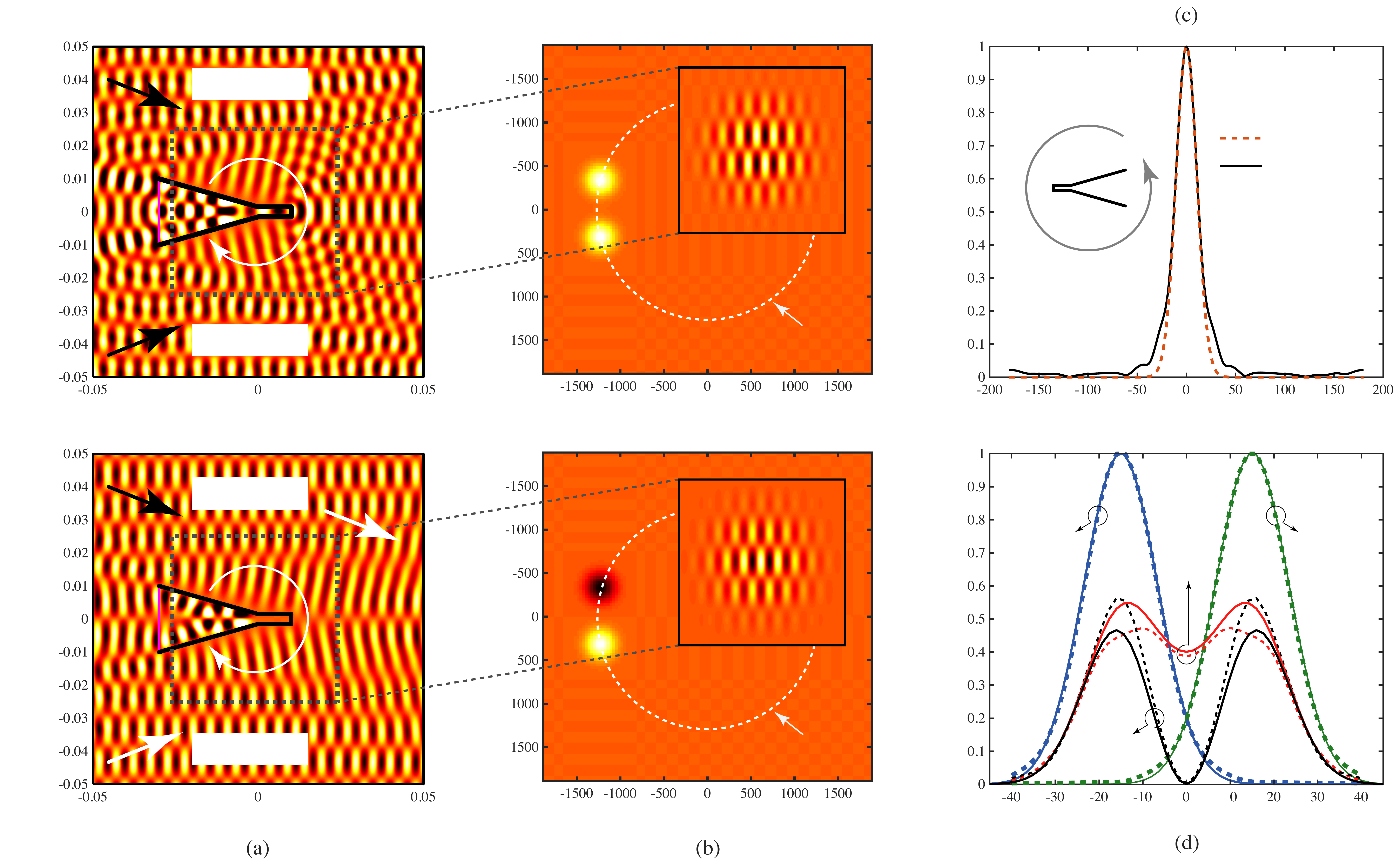}
     
     \put(17,32){\scriptsize{$x$~(m)}}
     \put(2.5,45){\rotatebox{90}{\scriptsize{$y$~(m)}}}
     \put(14.5,55.2){\scriptsize{$+15^\circ$ PW}}
     \put(14.5,36.9){\scriptsize{$-15^\circ$ PW}}
     
     \put(48,32){\scriptsize{$k_x$~(rad/m)}}
     \put(34,44){\rotatebox{90}{\scriptsize{$k_y$~(rad/m)}}}
     \put(53,43.5){\scriptsize{$x$~(m)}}
     \put(46.8,49.5){\rotatebox{90}{\scriptsize{$y$~(m)}}}
      \put(56,37){\scriptsize{$k_0$~circle}}
     
     \put(83,32){\scriptsize{$\phi$~(deg)}}
     \put(67,43){\rotatebox{90}{\scriptsize{$p(\phi)$ (norm.) }}}
     \put(90.5,49.5){\scriptsize{FEM-HFSS}}
     \put(90.5,51.5){\scriptsize{Gaussian fit}}
      \put(77.3,53.8){\scriptsize{$\phi$}}
     
     \put(17,3){\scriptsize{$x$~(m)}}
     \put(2.5,16){\rotatebox{90}{\scriptsize{$y$~(m)}}}
     \put(14.5,26){\scriptsize{$+15^\circ$ PW}}
     \put(14.5,7.6){\scriptsize{$-15^\circ$ PW}}
     
    \put(48,3){\scriptsize{$k_x$~(rad/m)}}
     \put(34,15){\rotatebox{90}{\scriptsize{$k_y$~(rad/m)}}}
     \put(53,14){\scriptsize{$x$~(m)}}
     \put(46.8,20){\rotatebox{90}{\scriptsize{$y$~(m)}}}
      \put(56,8){\scriptsize{$k_0$~circle}}
     
     \put(83,3){\scriptsize{$\theta$~(deg)}}
     \put(67,12){\rotatebox{90}{\scriptsize{Scene, $\mathcal{P}(\theta)$ (norm.)}}}
      \put(72,22.7){\tiny{$+15^\circ$ only}}
     \put(92,22.7){\tiny{$-15^\circ$ only}}
    \put(76.4,8){\tiny{out-of-phase}}
     \put(83,20.5){\tiny{in-phase}}
     
     \put(10,60.3){\color{amber}\footnotesize{\textsc{Horn Antenna (DFO)}}}
     \put(36,60.3){\color{amber}\footnotesize{\textsc{Structured Field Observation (SFO)}}}
     \put(0,38){\rotatebox{90}{\color{amber}\footnotesize{\textsc{in-Phase Plane-waves}}}}
     \put(0,7){\rotatebox{90}{\color{amber}\footnotesize{\textsc{out-of-Phase Plane-waves}}}}
    
     \end{overpic}
     \caption{Application of the proposed SFO method when an actual Horn antenna is used an observer to measure ideal plane-waves. a) Total fields computed using BEM method (POV $\theta = 180^\circ$, here) when two plane-waves illuminated the Horn under in-phase and out-of-phase conditions. b) The SFO method applied to the total fields without the horn, showing the $k_x-k_y$ fields, and the conditioned fields around the vicinity of the Horn in the inset. c) Gaussian fit of the FEM-HFSS computed radiation pattern of a typical horn. d) Comparison of the resulting SFO scene represented by $\mathcal{P}(\theta)$ (solid curves) with those obtained using BEM simulation of the Horn (dashed curves). The simulation parameters are: horn waveguide length 0.01~m, horn waveguide width of 0.001~m, horn flare length 0.03~m and horn aperture width of 0.02~m. }
     \label{fig:HornvsSFO}
 \end{figure*}

\subsection{Distributed Observer}

To obtain a more effective realization of an Observer we sample the field within a local region using a spatial weighting function, $w(\r)$, and then use a defined angular response of the observer to weight the incident fields. This method we will refer to as the Structured Field Observation (SFO). The weighting function is needed to sample the field in the region such that the field is attenuated at the edges to essentially zero but that the field at the center of the region is smoothly modulated. This weighting operation can be written as
\begin{align}\label{Eq:RectSample}
\psi^\text{obs.}(\r) = \psi^\text{tot.}(\r)w(\r),
\end{align}
where $\psi^\text{tot.}$ is the total fields in the vicinity of the observer, $w(\r)$ is the weighting function used to sample the fields, and  $\psi^\text{obs.}(\r)$ is the resulting conditioned fields. This sampling of the field allows for a spatial Fourier transform to be used to obtain the directional structure of the field in the local region around the Observer. We will use a Gaussian weighting function here, as it is simple and well suited to the spatial Fourier transform approach used. 

To illustrate this approach, in Fig. \ref{fig:ObsDef}(a), a complex field consisting of a Gaussian beam incident field (propagating left to right along the $x-$axis) and back scattered components off a number of objects. An Observer position is specified in the region and the weighting function is defined as a square area with an extent of $5\lambda$ around this point is defined as the observation area. In this region a high resolution field description is obtained using  propagation operators and the currents solved for on the surface of the objects. This field is presented in Fig.~\ref{fig:ObsDef}(b). To obtain the directional structure of this field, a 2D Fast Fourier Transform (FFT) can be performed as
\begin{align}\label{Eq:RectSampleFFT}
\tilde{\psi}^\text{obs.}(k_x, k_y)& = \mathcal{F}\{\psi^\text{tot.}(x,y)w(x,y)\} \\ \notag
&= \tilde{\psi}^\text{tot.}(k_x,k_y)\circledast\tilde{w}(k_x,k_y)
\end{align}
which is shown in Fig.~\ref{fig:ObsDef}(c), where the sign of the spatial frequencies $k_x$ and $k_y$ provides the directional information of the EM power flow. Although the directional components of the field can be seen in this plot, the field in Fig.~\ref{fig:ObsDef}(a) has effectively been sampled with a 2D pulse function and the resultant FFT has a Sinc function response convoluted ($\circledast$) with the actual field leading to undesirable artificial diffraction and a smeared response. 

This effect can be alleviated by sampling and conditioning the local field with a Gaussian function instead, with a width of $w_0 = 2.5\lambda$, for instance, given by
\begin{align}\label{Eq:GaussWeightFctn}
w(x,y) = \exp\left(\frac{x^2 + y^2}{w_0^2}\right),
\end{align}
where this width $w_0$ is considered one of the characteristic of the observer measurement capability. The resulting sampled field is presented in Fig.~\ref{fig:ObsDef}(d) and a 2D-FFT of the conditioned field is shown in \ref{fig:ObsDef}(e). The FFT clearly shows the scattered components of the field (left hand side of the plane and negative $k_x$) and the single incident wave on right hand side of the plane (positive $k_x$). Moreover, the spatial asymmetry of the scene with respect to the $x-$axis is also captured in the different strength of $\tilde{\psi}^\text{obs.}$ in the $k_x-k_y$ plane about the $k_x=0$ axis. Note the FFT was performed using zero padding to extend the size of the Observation region by 4 times to smooth the response. 

The field in Fig.~\ref{fig:ObsDef}(e) can be used to define an Observer with a POV and FOV and allow for a rendering of the scene from this position. A POV is first defined in the $x-y$ plane using an angle $\theta$ and the objective is to determine the total power measured by the observer, $\mathcal{P}(\theta)$, i.e. the scene, taking into account its directional measurement capability, $p(\phi)$ (akin to radiation patterns of electromagnetic antennas). The Fourier-transformed fields of Fig.~\ref{fig:ObsDef}(e) represents the power traveling through the physical region shown in Fig.~\ref{fig:ObsDef}(d) decomposed into plane-waves described by the wave-vectors $k_x$ and $k_y$ in the form of $e^{j\omega t + k_x x + k_yy}$. The angular receive function $p(\phi)$ of the observer is thus defined naturally in the $k_x-k_y$ domain, where it represents the power received by the observer when excited with a plane-wave incoming at an angle $\phi$. The receive function is assumed to be \textit{maximized} at $\phi = -\theta$ which represents a plane-wave incoming at normal incidence to the observer POV.  

 As the simulation is at a fixed $\omega$, all the waves will be present on the circle with a radius of $k_0$. In order to determine the total field incident on the observer from a particular direction POV, $\theta$, we first extract the fields present on the $k_0$ circle. The response function $p(\phi)$ is next orientated such that it points in the direction of the POV, i.e. maxima at $\phi = -\theta$. The fields on the circle are then weighted by the receive function and integrated over the circle. To obtain the total power $P(\theta)$, we then square the absolute magnitude of the field. This operation can be expressed as:
\begin{align}\label{Eq:PowerScan}
\mathcal{P}(\theta) = \left|\int_{\theta - \pi}^{\theta + \pi} p(\phi)\tilde{\psi}^\text{obs.}(\rho = k_0, \phi) d\phi \right|^2
\end{align}
To finally render a scene over the entire FOV, $\theta$ is simply scanned over the angular extent of interest and $P(\theta)$ determined. Such a response is shown in Fig. \ref{fig:ObsDef}(f), where $\mathcal{P}(\theta)$ is shown for the angular range of $\theta$ covering the second half of the $x-y$ region, for instance.

\begin{figure*}[htbp]
    \centering
      \begin{overpic}[grid=false, scale = 0.21]{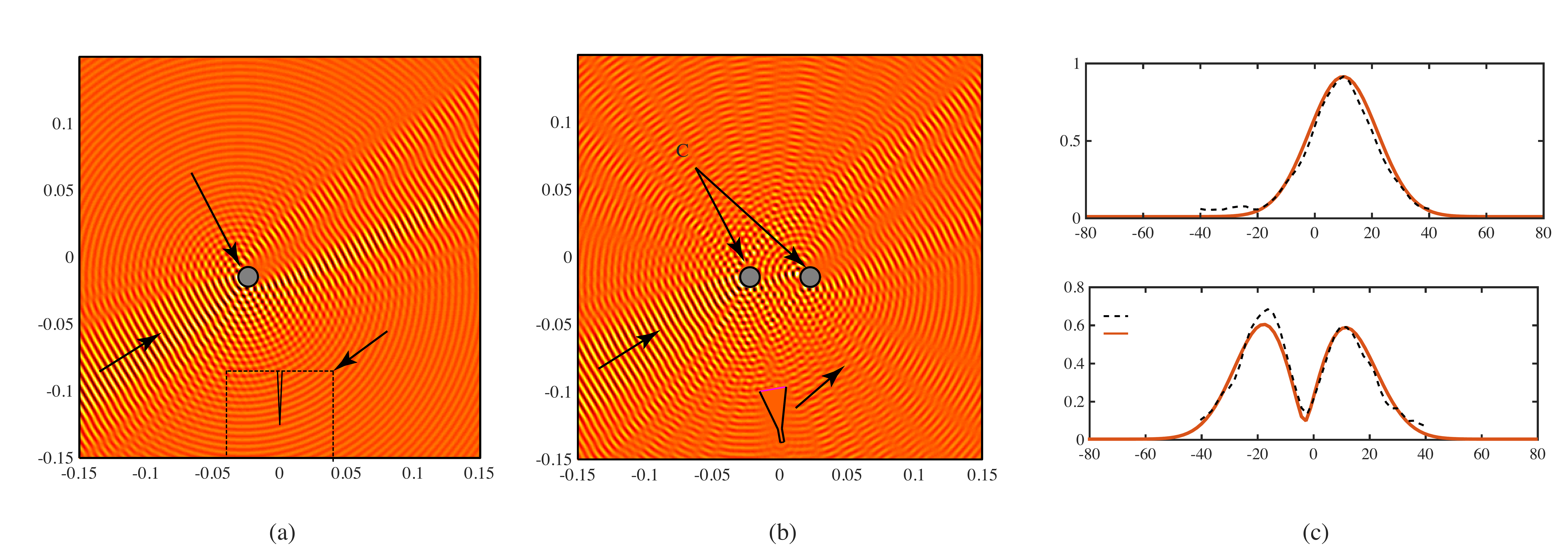} 
      
    \put(8,32.5){\footnotesize\shortstack{Single PEC Cylinder - SFO \\ Total Fields Re$\{E_z^\text{tot.}(x,y)\}$}}
    \put(16.5,3){\scriptsize{$x$~(m)}}
    \put(1.5,17){\rotatebox{90}{\scriptsize{$y$~(m)}}}
    \put(8,25){\scriptsize{PEC Cylinder}}
    \put(22,15){\scriptsize{Observer Region}}
    \put(7,16){\scriptsize{\shortstack{Incident\\Field}}}
    
    \put(40,32.5){\footnotesize\shortstack{Double PEC Cylinders - Horn \\ Total Fields Re$\{E_z^\text{tot.}(x,y)\}$}}
    \put(48.5,3){\scriptsize{$x$~(m)}}
    \put(32.5,17){\rotatebox{90}{\scriptsize{$y$~(m)}}}
    \put(40,25){\scriptsize{Observer Region}}
    \put(38,16){\scriptsize{\shortstack{Incident\\Field}}}
    \put(51,12){\scriptsize{Horn Antenna}}
    
    \put(82,4.5){\scriptsize{$\theta$ (deg)}}
    \put(65.5,10.5){\rotatebox{90}{\scriptsize{$\mathcal{P}(\theta)$}}}
    \put(73,28.6){\tiny{FEM-HFSS}}
   \put(73,27.6){\tiny{SFO}}
    
    \put(82,18){\scriptsize{$\theta$ (deg)}}
    \put(65.5,25.2){\rotatebox{90}{\scriptsize{$\mathcal{P}(\theta)$}}}
    \put(72.5,14.6){\tiny{FEM-HFSS}}
   \put(72.5,13.6){\tiny{SFO}}
    
\end{overpic}
    \caption{The scene computed using the SFO approach for two different cases of a single highly reflective cylindrical object and two interacting reflective cylinders, respectively, illuminated with identical Gaussian beams. a) Total field for a single cylindrical object using BEM. b) Two reflective circles with Horn as the observer. b) Comparison of the SFO method and the Horn based observation for the two cases. Simulation parameters are: cylinder radius 7.5~mm, locations $(-0.0225,-0.015)$~m and $(0.0225,-0.015)$~m, incident Gaussian beam of width 0.0375~m at $30^\circ$ with the observer located at $(0,-0.125)$~m.}
    \label{fig:hornvw}
\end{figure*}
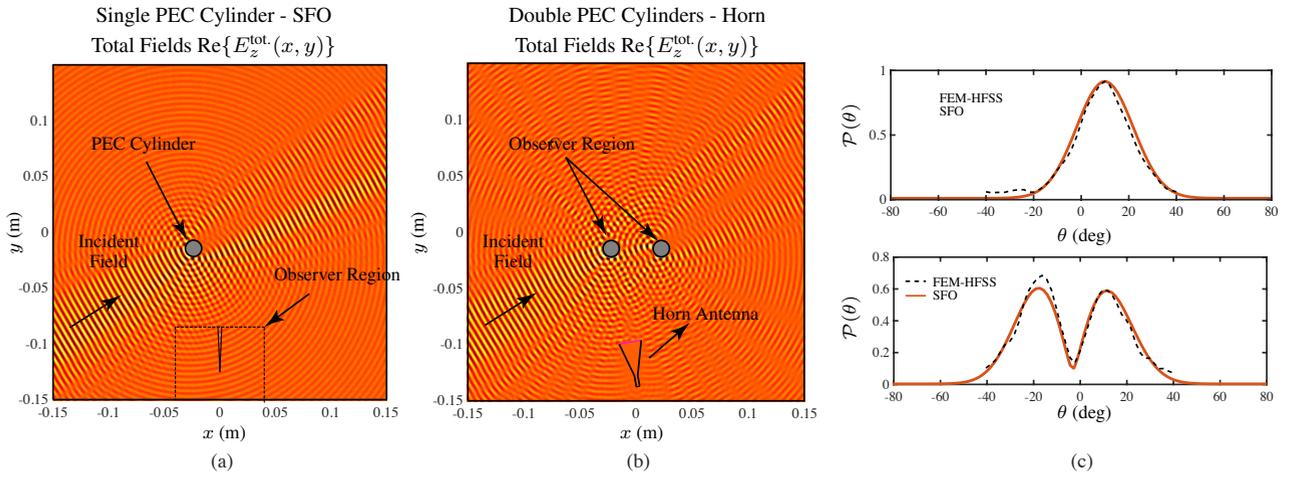

\subsection{Horn Antenna vs SFO}

A direct comparison of this distributed method of observation and an actual observation device can be made to test the efficacy of the above procedure in modeling the observer. For a radio frequency signals, an appropriate observing device is a \textit{Horn antenna}. A horn antenna is directional with an angular response defined by the input mode distribution at its front radiating aperture. Let us consider two obliquely propagating plane-waves ($\pm 15^\circ$) which are being received by a horn antenna as shown in Fig.~\ref{fig:HornvsSFO}(a). Since, the two waves form a standing wave along $y$, the exact placement of the horn is important. This can clearly be seen by setting the POV to $\theta = 180^\circ$ and using two specific conditions of in-phase and out-of-phase plane-waves at the aperture of the horn -- with a corresponding high and low reception of the waves as evident from the field distribution at the end of the waveguide section.

If such a horn antenna is to be used to render a scene, the horn must be rotated about $(0,0)$, for instance, and for every angle $\theta$, the full-wave simulation must be performed to compute the total power received by the horn to construct $\mathcal{P}(\theta)$. This will correspond to a Direct Field Observation (DFO) with the horn in place. While it provides accurate and rigorous description of the scattered fields around the horn, it is computationally expensive, and lacks deeper physical insight. Let us apply the Structure Field Observation (SFO) to reconstruct the scene. In this case, the horn is \textit{removed from the scene}, and the total fields are computed. In the region of interest around the horn, the field is conditioned by the spatial Gaussian function and the 2D-FFT taken, as shown in Fig.~\ref{fig:HornvsSFO}(b) for the two cases of in-phase and out-of-phase conditions. For each case, the two plane-waves are clearly identified lying on the $k_0$ circle, with correct phase relationship between the two, evident in the symmetric and antisymmetric field magnitudes. Using this $k_x-k_y$ domain field representation, the scene can now be rendered following the methodology of Fig.~\ref{fig:ObsDef}.

Before proceeding further, the angular receive function, $p(\phi)$ of the horn must be determined. Fig.~\ref{fig:HornvsSFO}(c) shows the co-polarized radiation pattern (power gain) of the horn obtained using HFSS, showing a 3-dB beamwidth of $18^\circ$ at 60~GHz. For convenience, this receive function can be approximated by a Gaussian function, as shown in Fig.~\ref{fig:HornvsSFO}(c). Clearly, the Gaussian function approximates the horn pattern well, particularly the  main beam, while failing to capture the side-lobes. Although, in principle the FEM-HFSS pattern may be directly used as the $p(\phi)$, the Gaussian function represents a simple and a generic form of the receive function that may be applicable to wide variety of observers. Next, the scene is rendered using \eqref{Eq:PowerScan} with the assumed Gaussian receive function, and the computed power scan is shown in Fig.~\ref{fig:HornvsSFO}(d). As a sanity check, the SFO scene of determined using single plane-waves, one at a time were compared with those computed using FEM-HFSS (solid vs dashed curves). A near-perfect match is obtained between the two, thus providing confidence that the SFO captures the basic behavior of a directionally sensitive observer. Next, both plane-waves are present and two computed scenes are compared with those using full-wave HFSS in Fig.~\ref{fig:HornvsSFO}(d) for cases of in and out-of-phase plane-wave conditions, respectively. A good agreement is obtained between the two, where the two plane-waves are seen to be resolved with correct magnitudes at $\theta=0^\circ$. While the overall power distribution is correctly captured by the SFO, the discrepancies with the full-wave results can be attributed to the actual scattering by the horn which are naturally not present in the SFO method. This is particularly attractive, since the use of SFO is very efficient as it provides the complete angular response with a single simulation and does not require a set of simulations as does a direct simulation of a Horn in either HFSS or using the IE method. 

To further confirm the appropriateness of the use of the SFO, Fig.~\ref{fig:hornvw} shows a comparison between direct Horn simulations using the IE framework and the SFO for two different scenes with more complexity. These two examples scenes are: 1) a single highly reflective cylindrical object; and 2) two interacting reflective cylinders (Fig.~\ref{fig:hornvw}(a) and \ref{fig:hornvw}(b) respectively). A comparison of rendering the two scenes is presented in Fig.~\ref{fig:hornvw}(c). A very good agreement is observed between the brute force IE method and the SFO, thereby justifying the usability of the SFO for constructing an Observer scene. Next, several examples will be presented using the proposed IE-GSTC infrastructure and the SFO model of the observer, to demonstrate the camouflaging and illusion forming capabilities of a closed metasurface structure.

\section{Examples: Illusion \& Camouflaging}\label{Sec-V}

To illustrate the synthesis procedure for metasurface based illusions and object camouflaging, we will first consider some basic examples and then progress to more complicated cases involving merged illusions and dynamic illusions. The simulation setup follows the illustration of Fig.~\ref{Fig:SimSetup}, in the $x-y$ plane, and the frequency of operation is chosen to be $60$~GHz ($\lambda = 5$~mm). The surface meshing is set to $\lambda/10$ based on proper convergence. The region over which fields were plotted was 40$\lambda$ in both the $x$ and $y$ directions. The objects from which the illusions were created are of the order of 10's of $\lambda$ is size, with a variety of shapes chosen to illustrate the methods ability to capture the full wave diffraction present. Objects with sharp corners produce large diffractive effects, but smoother objects can produce lens like effects. Although these objects were kept moderate in size to illustrate effects. It should be noted that for the IE-BEM method the fields are plotted post-simulation and the region to be plotted is an arbitrary choice.  Only the scalar electric and magnetic surface susceptibilities are used which were found sufficient to create the illusions. More complex susceptibility distributions utilizing other tensorial elements may be used to satisfy possible practical constraints such as lossless-ness and reciprocity as mentioned in Sec.~III. As will be seen, all the surface susceptibilities are found to be passive and physical with appropriate choices of the illumination fields.

\subsection{Simple Illusions}

\begin{figure}[htbp]
    \centering
          \begin{overpic}[grid=false, scale = 0.21]{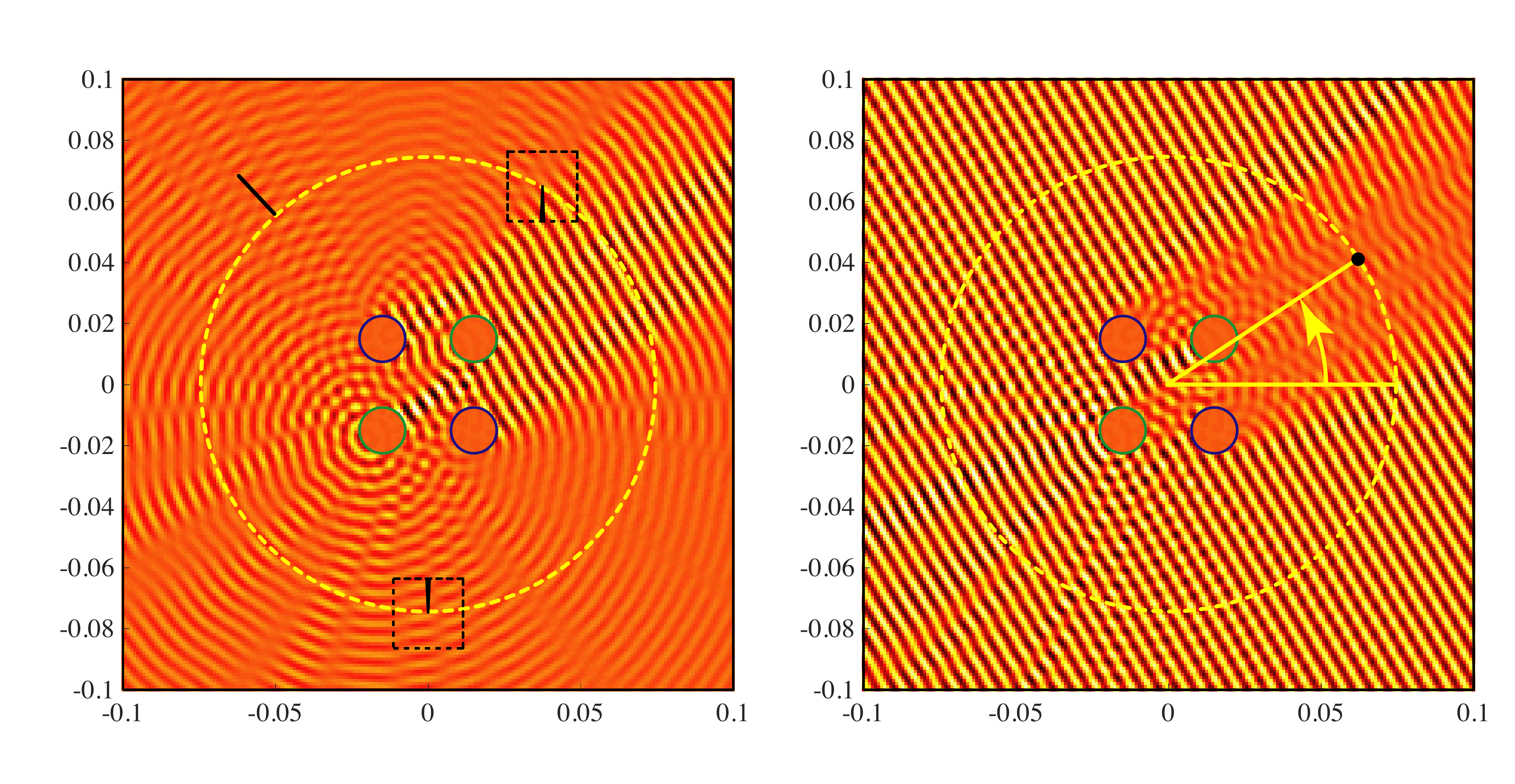} 
          
    \put(26,1){\scriptsize{$x$~(m)}}
    \put(1,25){\rotatebox{90}{\scriptsize{$y$~(m)}}}
     \put(10,48){\scriptsize{Scattered Fields, Re$\{E_z(x,y)\}$}} \put(32,42.5){\scriptsize{POV \#1}}
      \put(25,15){\scriptsize{POV \#2}}
    \put(9,41){\scriptsize{\shortstack{Observer \\Scan}}}
      
    \put(76,1){\scriptsize{$x$~(m)}}
    \put(50,25){\rotatebox{90}{\scriptsize{$y$~(m)}}}
      \put(62,48){\scriptsize{Total Fields, Re$\{E_z(x,y)\}$}}
    \put(89,29){\scriptsize{$\theta$}}
     \put(89,36){\scriptsize{$(r_0, \theta)$}}
    
\end{overpic}
    \caption{The scene computed using the SFO approach for two highly reflective cylindrical object and two interacting reflective cylinders, respectively, illuminated with a uniform plane wave. The simulation parameters are: incident plane-wave at $30^\circ$, four cylinders of $ r = 0.0075$~m located at  $(\pm 0.015,\pm0.015)$ with reflectivities $R_1 = R_4 = 0.3$, $R_2 = R_3 = 0.9$ and transmission $T= 0$; two observers located at $(0.0375,0.065)$ and $(0.0,-0.075)$. The radius of the circle for Direct Field Observation (DFO) is $0.075$~m.}
    \label{Fig:SimpleIllusion_Ref}
\end{figure}

\begin{figure*}[htbp]
    \centering
	 \begin{overpic}[grid=false, scale = 0.24]{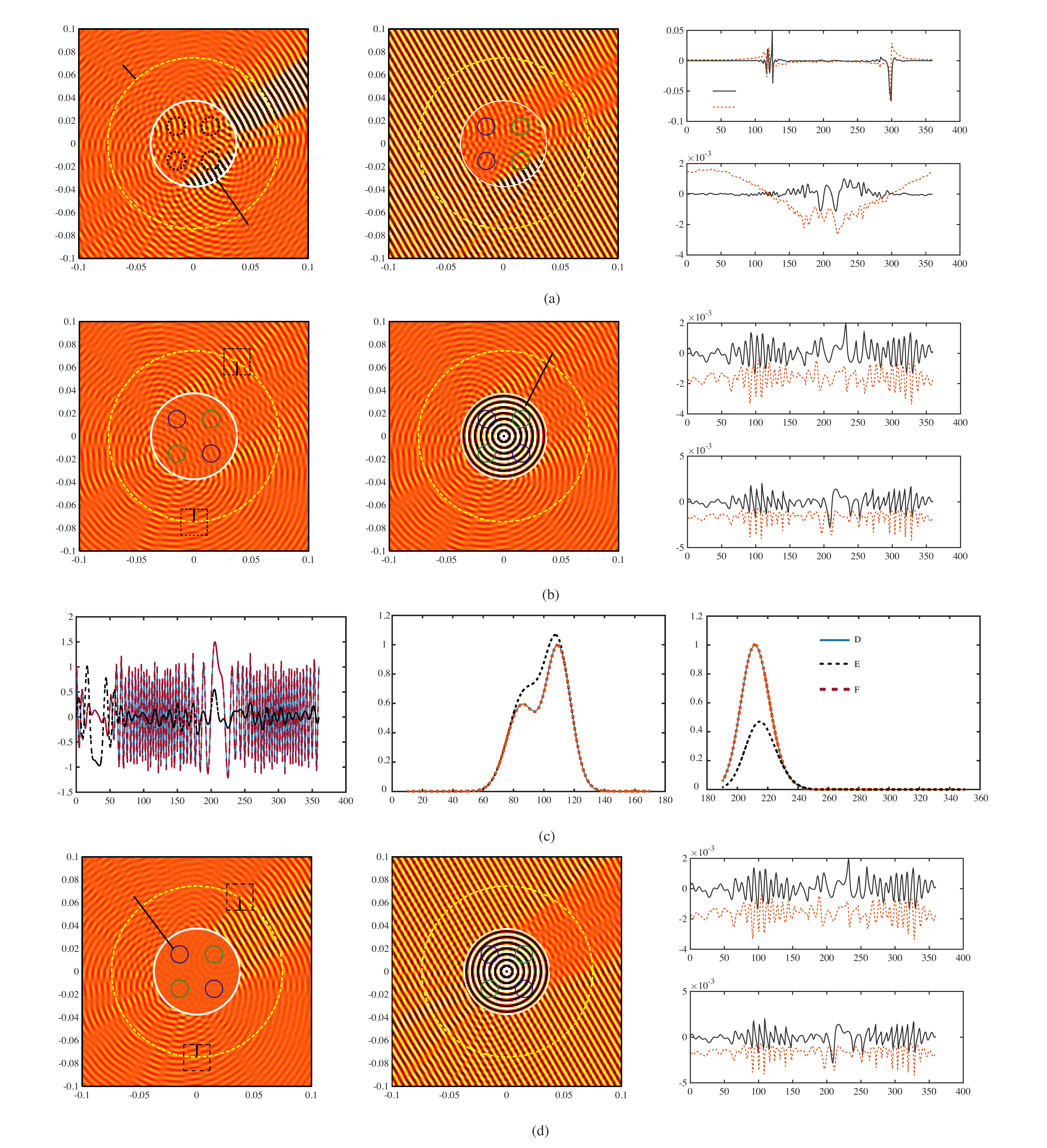} 
	 
    \put(15,75){\scriptsize{$x$~(m)}}
    \put(3,86.5){\rotatebox{90}{\scriptsize{$y$~(m)}}}
    \put(8,94.5){\scriptsize{\shortstack{Observer \\Scan}}}
    \put(19,77.7){\scriptsize{\shortstack{Metasurface \\Hologram}}}
    \put(7,98){\scriptsize{Scattered Fields, Re$\{E_z(x,y)\}$}}
    
    \put(42.2,75){\scriptsize{$x$~(m)}}
    \put(30,86.5){\rotatebox{90}{\scriptsize{$y$~(m)}}}
   \put(35,98){\scriptsize{Total Fields, Re$\{E_z(x,y)\}$}}
    
    \put(70,87){\scriptsize{$\theta$~(deg)}}
    \put(56,90){\rotatebox{90}{\scriptsize{$\chi_\text{mm}(r_m, \theta)$}}}
        \put(64.5,91.8){\tiny{Re$\{\cdot\}$}}
        \put(64.5,90.5){\tiny{Im$\{\cdot\}$}}
    
    \put(71,75){\scriptsize{$\theta$~(deg)}}
    \put(56,79){\rotatebox{90}{\scriptsize{$\chi_\text{ee}(r_m, \theta)$}}}

    \put(15,49.5){\scriptsize{$x$~(m)}}
    \put(3,61){\rotatebox{90}{\scriptsize{$y$~(m)}}}
    \put(18.5,70){\scriptsize{POV \#2}}
    \put(14.5,56.2){\scriptsize{POV \#1}}
    
    \put(42.2,49.5){\scriptsize{$x$~(m)}}
    \put(30,61){\rotatebox{90}{\scriptsize{$y$~(m)}}}
      \put(44,70){\scriptsize{Internal Illumination}}
    
    \put(70,61){\scriptsize{$\theta$~(deg)}}
    \put(57,64){\rotatebox{90}{\scriptsize{$\chi_\text{mm}(r_m, \theta)$}}}
    
    \put(71,49.5){\scriptsize{$\theta$~(deg)}}
    \put(57,53){\rotatebox{90}{\scriptsize{$\chi_\text{ee}(r_m, \theta)$}}}

    \put(15,3){\scriptsize{$x$~(m)}}
    \put(3,14){\rotatebox{90}{\scriptsize{$y$~(m)}}}
        \put(18.5,23.3){\scriptsize{POV \#2}}
    \put(14.5,9.5){\scriptsize{POV \#1}}
    \put(8,22.2){\scriptsize{\shortstack{Virtual \\Objects}}}
    
    \put(42.2,3){\scriptsize{$x$~(m)}}
    \put(30,14){\rotatebox{90}{\scriptsize{$y$~(m)}}}
    
    \put(70,14.8){\scriptsize{$\theta$~(deg)}}
    \put(57,18){\rotatebox{90}{\scriptsize{$\chi_\text{mm}(r_m, \theta)$}}}
    
    \put(71,2.8){\scriptsize{$\theta$~(deg)}}
    \put(57,6.6){\rotatebox{90}{\scriptsize{$\chi_\text{ee}(r_m, \theta)$}}}

     \put(85,97){\rotatebox{270}{\small{\color{amber}\boxed{\shortstack{\textsc{External} \\ \textsc{Illumination Only}}}}}}
     \put(85,72){\rotatebox{270}{\small{\color{cyan}\boxed{\shortstack{\textsc{Internal} \\ \textsc{Illumination Only}}}}}}
     \put(85,26){\rotatebox{270}{\small{\color{ceruleanblue}\boxed{\shortstack{\textsc{Internal \& External} \\ \textsc{Illumination}}}}}}

      \put(16,28.5){\scriptsize{$\theta$~(deg)}}
        \put(3,35){\rotatebox{90}{\scriptsize{Re$\{E_z(r_0, \theta)\}$}}}
      
      \put(44,28.5){\scriptsize{$\theta$~(deg)}}
        \put(31,35){\rotatebox{90}{\scriptsize{$\mathcal{P}(\theta)$}}}
        \put(36,44){\scriptsize{POV \#2}}
      
      \put(72,28.5){\scriptsize{$\theta$~(deg)}}
        \put(58.6,37){\rotatebox{90}{\scriptsize{$\mathcal{P}(\theta)$}}}
        \put(80,32){\scriptsize{POV \#1}}
         \put(75,44){\tiny{reference scene, Fig.~\ref{Fig:SimpleIllusion_Ref}}}
         \put(75,41){\tiny{\shortstack{internal \\ illumination only}}}
        \put(75,38.5){\tiny{\shortstack{internal and external \\ illumination}}}

    \end{overpic}
    \caption{Metasurface hologram recreating the reference case of Fig.~\ref{Fig:SimpleIllusion_Ref}. Total and scattered fields, along with synthesize surface susceptibilities using a) external and b) internal illuminations only. c) Comparison of the observation fields and the scenes constructed at two POVs. d) Total and scattered fields, along with synthesize surface susceptibilities using both external and internal illuminations. The external illumination is the same at the incident fields of Fig.~\ref{Fig:SimpleIllusion_Ref} and the radius of the metasurface hologram is $0.0375$~m.}
    \label{Fig:Simple_Illusion_Circle}
\end{figure*}

Let us consider a reference case consisting of 4 circular cylinders made up of two different materials (PEC and dielectric), which are excited with a wide Gaussian beam. The scattered and the total fields produced as a result are shown in Fig.~\ref{Fig:SimpleIllusion_Ref}. This collection of objects, produce a large shadow behind them. An observation circle is defined at $r_0$ for DFO. In addition, two observers are placed in the upper and lower halves, respectively, to construct the SFO. We now wish to reconstruct the same fields and scenes by removing the objects and replacing them with a single metasurface hologram, as shown in Fig.~\ref{Fig:Simple_Illusion_Circle}.

\begin{figure*}[htbp]
    \centering
    	 \begin{overpic}[grid=false, scale = 0.27]{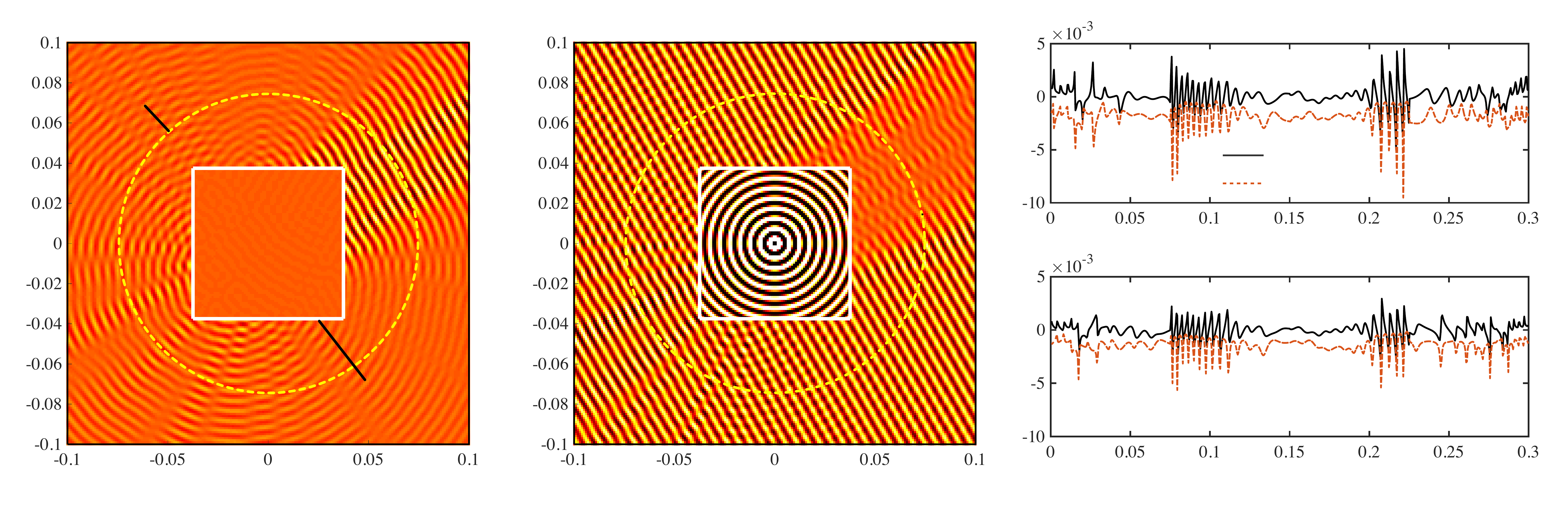} 
    	 
    \put(16,1.2){\scriptsize{$x$~(m)}}
    \put(0.5,15.3){\rotatebox{90}{\scriptsize{$y$~(m)}}}
        \put(5.5,26){\scriptsize{\shortstack{Observer \\Scan}}}
      \put(21,5){\scriptsize{\shortstack{Metasurface \\Hologram}}}
         \put(8,31){\scriptsize{Scattered Fields, Re$\{E_z(x,y)\}$}} 
         
    \put(48.5,1.2){\scriptsize{$x$~(m)}}
    \put(32.5,15.3){\rotatebox{90}{\scriptsize{$y$~(m)}}}
    \put(42,31){\scriptsize{Total Fields, Re$\{E_z(x,y)\}$}}
    
     \put(78,16){\scriptsize{distance~(mm)}}
    \put(64,21){\scriptsize\rotatebox{90}{$\chi_\text{mm}(r_m, \phi)$}}
        \put(81,22){\tiny{Re$\{\cdot\}$}}
        \put(81,20.3){\tiny{Im$\{\cdot\}$}}
        
     \put(78,1.2){\scriptsize{distance~(mm)}}
    \put(64,6){\scriptsize\rotatebox{90}{$\chi_\text{ee}(r_m, \phi)$}}
    
    %  \psfrag{H}[c][c][0.8]{$\theta$~(deg)}
    %  \psfrag{P}[c][c][0.8]{$\mathcal{P}(\theta)$}
    %  \psfrag{Q}[c][c][0.8]{Re$\{E_z(r_0, \theta)\}$}
    
\end{overpic}
    \caption{Metasurface recreating the reference case of Fig.~\ref{Fig:SimpleIllusion_Ref} using a square shaped hologram (of side 7.5~cm), showing the total and scattered fields, along with synthesized surface susceptibilities using both external and internal illuminations.}
    \label{Fig:Simple_Illusion_Square}
\end{figure*}

Let us consider a natural choice of incident fields as the illumination fields \textit{external} to the metasurface. Fig.~\ref{Fig:Simple_Illusion_Circle}(a) shows the computed scattered and total fields everywhere in the region around the surface. Comparison with the reference fields of Fig.~\ref{Fig:SimpleIllusion_Ref} shows a clear and a large discrepancy in the reconstructed fields, where the metasurface synthesis essentially fails. The method produces susceptibilities that are large, active and disjointed, where the simulation becomes very sensitive to illumination intensity. This poor match may be attributed to a non-uniform one-sided illumination of surface, where the surface fails to re-create the fields to ``shape'' the shadow to match the illusion.

Noting that the previous simulation failed due to non-uniform illumination a natural modification is to use a uniform internal illumination. We thus chose an internal illumination that is a radially propagating wave from center of the metasurface hologram with $|E_z|= 2$ at the surface. The resulting total and scattered fields are shown in Fig.~\ref{Fig:Simple_Illusion_Circle}(b), where this time the illusions is better formed and reproduces the reflected scattered fields in the reference scene well. The corresponding surface susceptibilities are well-behaved and are passive as also shown in Fig.~\ref{Fig:Simple_Illusion_Circle}(b). However, the illusion can not reproduce the \textit{total fields} throughout the scene as the fields are restricted to outwardly propagating waves. Only within the shadow does it reasonably produce the total fields. This is clearly evident in the DFO and SFO results shown in Fig.~\ref{Fig:Simple_Illusion_Circle}(c), compared to the reference case. While the POV\#2 lying in the shadow region (with mostly scattered fields) is reconstructing the scene well to some extent, the POV\#1 fails and measures significantly reduced fields. We thus conclude that the internal illumination only is also not sufficient to correctly produce the illusion.

Thus the only way to recreate the total fields throughout the region is to have the incident field also present in the illusion scene, so that we have both an internal illumination (radially propagating internal wave) and an external illumination identical to the original incident fields. Fig.~\ref{Fig:Simple_Illusion_Circle}(d) shows the resulting scattered and total fields along with the synthesized surface susceptibilities for this case. This time a perfect reconstruction of the reference fields of Fig.~\ref{Fig:SimpleIllusion_Ref} is observed proving that its critical to have both internal and external illumination fields to be present. Internal illumination creates the fields on the right side of the illusion surface, as in this region, it only needs to create the outward propagating scattered fields that form the shadow. On the left side it can not create the total field on the surface due to the need to create inwardly propagating waves. Adding an external illumination assists in recreating this part of the fields and the match between the reference and reconstructed fields is prefect. Consequently, the DFO and SFO accurately agree with the reference scene as shown in Fig.~\ref{Fig:Simple_Illusion_Circle}(c).

Finally, to show the versatility of the method and the metasurface hologram, the same reference scene of Fig.~\ref{Fig:SimpleIllusion_Ref} can be recreated using a shape different than a circular surface. Fig.~\ref{Fig:Simple_Illusion_Square} shows one such example, where a rectangular shaped metasurface hologram is used. As expected, the fields recreated perfectly similar to that of Fig.~\ref{Fig:Simple_Illusion_Circle}, except this time, the surface susceptibilities across the surface are different, and showing larger spatial variation around the two corner regions of the square.

\subsection{Angle-dependent Metasurface Illusions}

\begin{figure*}[htbp]
    \centering
        	 \begin{overpic}[grid=false, scale = 0.2]{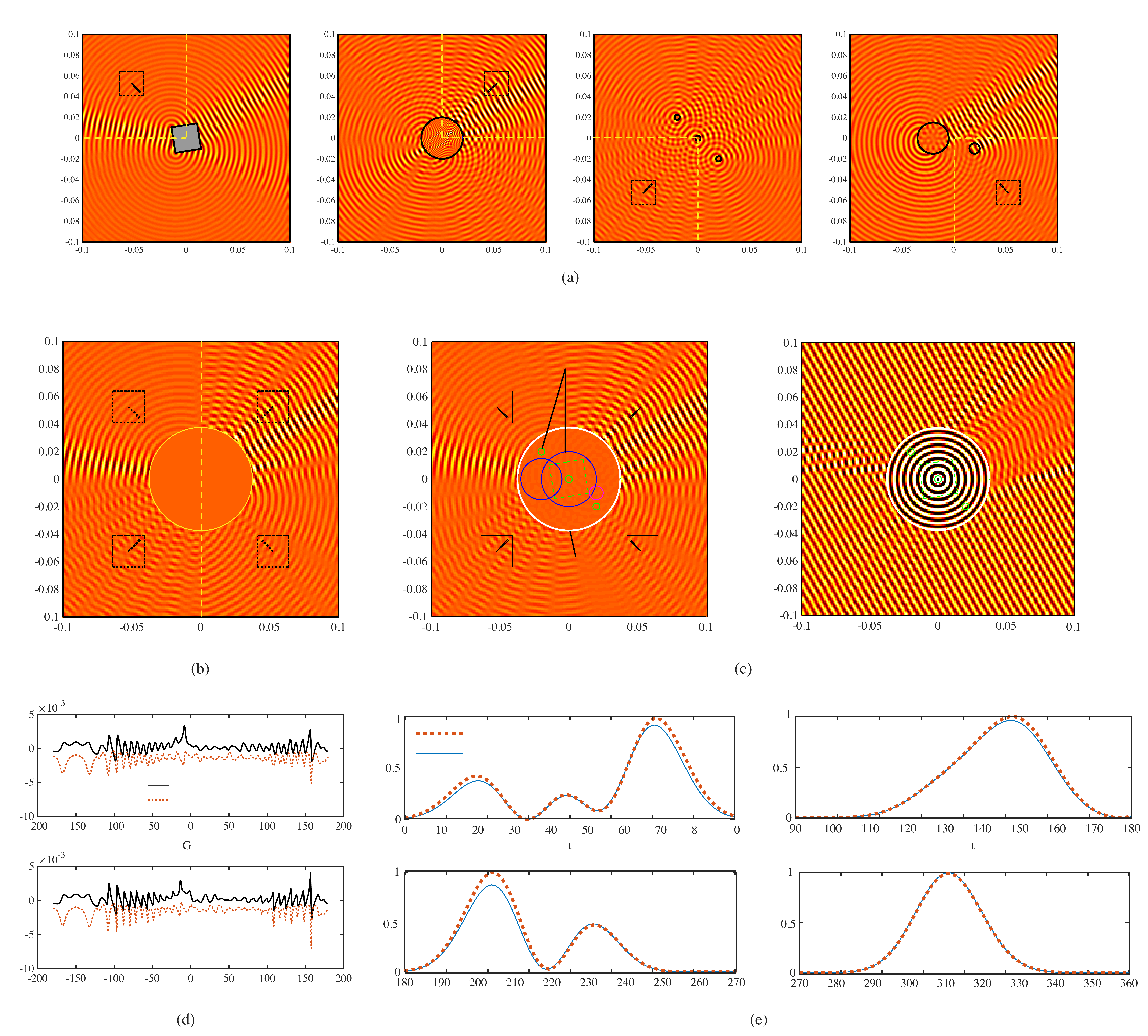} 
        	 
    \put(14.5,67){\scriptsize{$x$~(m)}}
    \put(3,76.5){\rotatebox{90}{\scriptsize{$y$~(m)}}}
        \put(9.5,84.5){\color{white}\scriptsize{POV \#2}}
           \put(9,88){\scriptsize{\shortstack{Scattered Fields\\ Re$\{E_z(x,y)\}$:  Square}}}
    
     \put(37,67){\scriptsize{$x$~(m)}}
    \put(26,76.5){\rotatebox{90}{\scriptsize{$y$~(m)}}}
    \put(41.5,84.5){\color{white}\scriptsize{POV \#1}}
        \put(28,88){\scriptsize{\shortstack{Scattered Fields\\ Re$\{E_z(x,y)\}$: Dielectric Cylinder}}}
    
     \put(60,67){\scriptsize{$x$~(m)}}
    \put(48,76.5){\rotatebox{90}{\scriptsize{$y$~(m)}}}
        \put(53.5,71){\color{white}\scriptsize{POV \#3}}
            \put(51,88){\scriptsize{\shortstack{Scattered Fields\\ Re$\{E_z(x,y)\}$: Three Cylinders}}}
    
     \put(81,67){\scriptsize{$x$~(m)}}
    \put(70.7,76.5){\rotatebox{90}{\scriptsize{$y$~(m)}}}
        \put(85.5,71){\color{white}\scriptsize{POV \#4}}
            \put(75,88){\scriptsize{\shortstack{Scattered Fields\\ Re$\{E_z(x,y)\}$: Two Cylinders}}}

     \put(15,34){\scriptsize{$x$~(m)}}
    \put(0.5,47){\rotatebox{90}{\scriptsize{$y$~(m)}}}
        \put(9.5,57){\color{white}\scriptsize{POV \#2}}
    \put(22,57){\color{white}\scriptsize{POV \#1}}
    \put(9.5,39.5){\color{white}\scriptsize{POV \#3}}
    \put(22,39.5){\color{white}\scriptsize{POV \#4}}
        \put(10,61.5){\scriptsize{\shortstack{Merged Fields\\ Re$\{E_z(x,y)\}$:  All Objects}}}
    
     \put(49,34){\scriptsize{$x$~(m)}}
    \put(33.5,47){\rotatebox{90}{\scriptsize{$y$~(m)}}}
        \put(41.5,57){\color{white}\scriptsize{POV \#2}}
    \put(54,57){\color{white}\scriptsize{POV \#1}}
    \put(41.5,39.5){\color{white}\color{white}\scriptsize{POV \#3}}
    \put(54,39.5){\color{white}\scriptsize{POV \#4}}
        \put(38,61.5){\scriptsize{\shortstack{Scattered Fields\\ Re$\{E_z(x,y)\}$:  Metasurface Hologram}}}
       \put(45,59){\scriptsize{\color{white} \shortstack{Virtual Objects}}}
    \put(47,39){\scriptsize{\color{white}\shortstack{Metasurface\\Hologram}}}
    
     \put(80,34){\scriptsize{$x$~(m)}}
    \put(65.5,47){\rotatebox{90}{\scriptsize{$y$~(m)}}}
        	    \put(71,61.5){\scriptsize{\shortstack{Total Fields\\ Re$\{E_z(x,y)\}$:  Metasurface Hologram}}}

    \put(14,16){\scriptsize{$\theta$~(deg)}}
    \put(0.5,20){\rotatebox{90}{\scriptsize{$\chi_\text{mm}(r_m, \theta)$}}}
    \put(15,21.6){\tiny{Re$\{\cdot\}$}}
   \put(15,20.2){\tiny{Im$\{\cdot\}$}}
   
 \put(48,16){\scriptsize{$\theta$~(deg)}}
   \put(32,20){\rotatebox{90}{\scriptsize{$\mathcal{P}^\text{POV\#1}(\theta)$}}}
          \put(40.5,26){\tiny{Reference Scene}}
      \put(40.5,24.5){\tiny{Metasurface SFO}}
      
       \put(83,16){\scriptsize{$\theta$~(deg)}}
     \put(65.8,20){\rotatebox{90}{\scriptsize{$\mathcal{P}^\text{POV\#2}(\theta)$}}}

 \put(14,3){\scriptsize{$\theta$~(deg)}}
    \put(0.5,6.5){\rotatebox{90}{\scriptsize{$\chi_\text{ee}(r_m, \theta)$}}}

 \put(48,3){\scriptsize{$\theta$~(deg)}}
      \put(32,6.5){\rotatebox{90}{\scriptsize{$\mathcal{P}^\text{POV\#3}(\theta)$}}}
      
 \put(83,3){\scriptsize{$\theta$~(deg)}}
       \put(65.8,6.5){\rotatebox{90}{\scriptsize{$\mathcal{P}^\text{POV\#4}(\theta)$}}}
   
   \end{overpic}
    \caption{An angle dependent illusion created by a metasurface hologram by merging four different scenes. a) Reference scattered fields corresponding to 4 different object configurations. b) Reference scattered fields obtained by merging scattered fields of each case taken from different quadrants. c) Recreated total and scattered fields using a metasurface hologram. d) The synthesized metasurface surface susceptibilities and e) resulting scenes created by four observers located in each quadrant. The simulation parameters are: PEC Square of 2.5~cm side length rotated by 10$^\circ$; Two cylinders located at $(-0.02,0)$~m and $(0.02,-0.01)$~m with radii of $\{0.015,~0.00335\}$~m, with reflectance and transmission of $\{0.9,~0.1\}$ and $\{0.1,~0.9\}$, respectively;  Dielectric Cylinder of radius $0.04$~m, and refractive index $3$ located at the origin $(0,0)$; Three small PEC circles of radius $r = 0.0025$~m located at $(-0.01,0.01)$, $(0.0,0.0)$, and $(0.01,-0.01)$, respectively; Four observers located at $(\pm0.0525,\pm0.0525)$ and the radius of the metasurface hologram is 0.075~m.}
    \label{Fig:MergedIlusion}
\end{figure*}

In the previous example, an observer moving about the metasurface (and thus around the virtual object), will measure identical fields compared to that of the reference scene, and thus perceive the same object from all angles of observations. However, the metasurface hologram maybe engineered to reproduce almost arbitrary fields around it so that different illusions may be projected and the observer perceives different objects when viewing the surface from different directions. Such a design capability is illustrated in Fig.~\ref{Fig:MergedIlusion}, where four different object configurations are shown in Fig.~\ref{Fig:MergedIlusion}(a), with variety of objects producing significantly different scattered fields. We now wish to a synthesize a metasurface hologram that can recreate the total fields produced by these four configurations in four quadrants simultaneously. To measure the effectiveness of this illusion, four observers with their own POVs are placed across all quadrants.

The first step in this synthesis is to spatially filter each of the reference configurations and combine them into a single scene. A 2D Sigmoid function with transition width $w = 0.5\lambda$ used here as a spatial filter, for instance, which is given by 
\begin{align*}
    f(x,y) = \left(\frac{1}{1+e^{-x/w}}\frac{1}{1+e^{-y/w}}\right).
\end{align*}
\noindent The resulting combined scattered fields are shown in Fig.~\ref{Fig:MergedIlusion}(b), along with an intended location of the metasurface hologram. Next, the metasurface is synthesized with both internal and external illumination along with the desired scattered fields of Fig.~\ref{Fig:MergedIlusion}(b). The resulting total and scattered fields are shown in Fig.~\ref{Fig:MergedIlusion}(c). An excellent recreation of the fields is observed all around the metasurface, whose surface susceptibilities are shown in Fig.~\ref{Fig:MergedIlusion}(d). The scene measured by the four observers in each quadrant is shown further in Fig.~\ref{Fig:MergedIlusion}(e) in comparison with the reference scene, which are computed using the SFO method. A very good agreement is observed in all cases. The slight discrepancies are due to the finite and smooth field transitions introduced by the Sigmoid function when moving from one quadrant to the other. The four reconstructed scenes are thus specific to the observer in their respective quadrants, and an observer moving around the metasurface hologram will sequential perceive these four object configurations, and experience an angle dependent illusion.

\begin{figure*}[!t]
    \centering
            	 \begin{overpic}[grid=false, scale = 0.29]{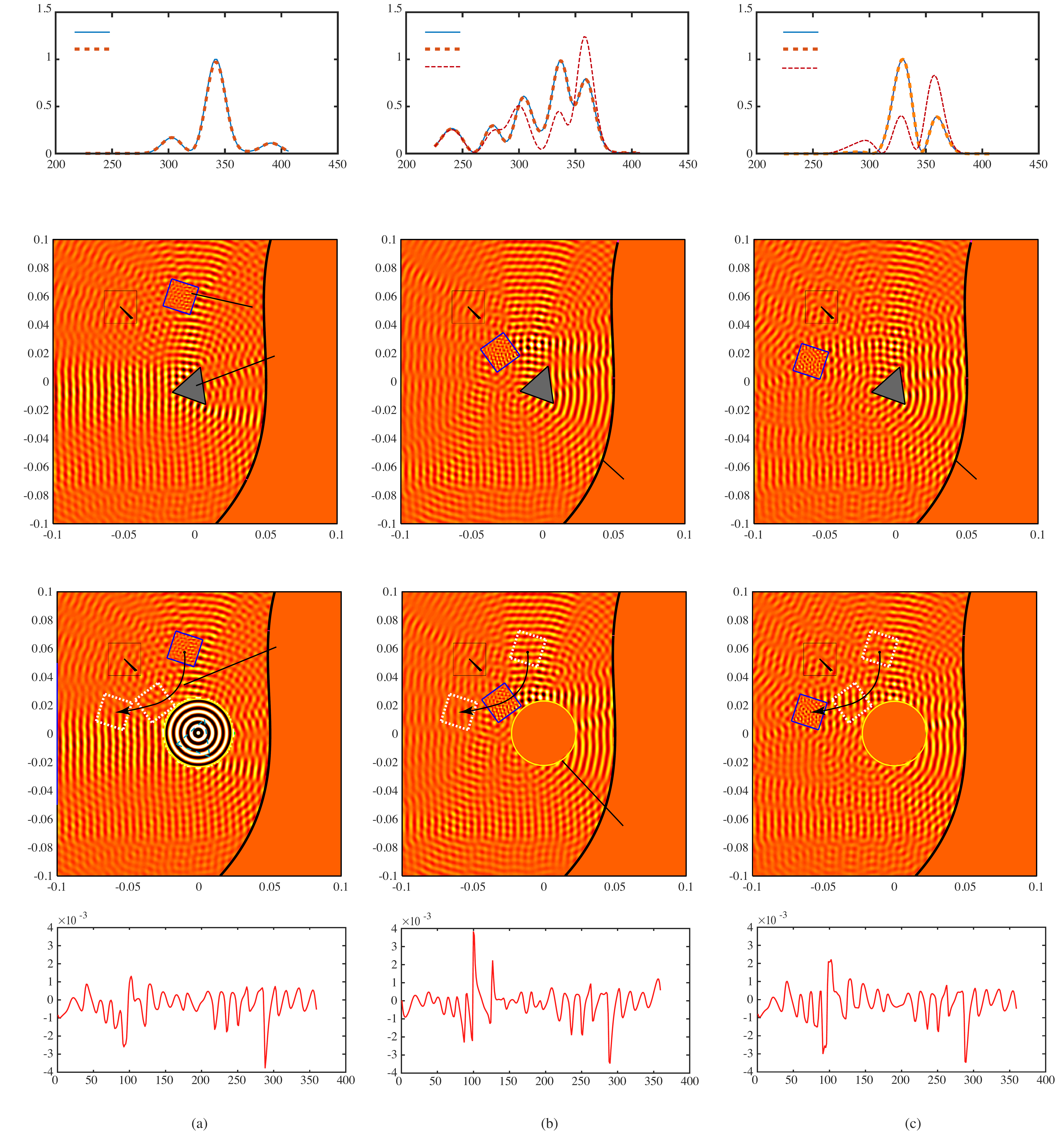} 
            	 
        \put(15,84){\scriptsize{$\theta$~(deg)}}
    \put(1.5,91){\rotatebox{90}{\scriptsize{$\mathcal{P}(\theta, t_0)$}}}
        \put(10.1,97){\tiny{Reference Scene}}
      \put(10.1,95.5){\tiny{Dynamic SFO}}
    
    \put(45,84){\scriptsize{$\theta$~(deg)}}
    \put(32,91){\rotatebox{90}{\scriptsize{$\mathcal{P}(\theta, t_1)$}}}
        \put(41,97){\tiny{Reference Scene}}
      \put(41,95.5){\tiny{Dynamic SFO}}   
     \put(41,93.8){\tiny{Static SFO} }
    
    \put(78,84){\scriptsize{$\theta$~(deg)}}
    \put(62.5,91){\rotatebox{90}{\scriptsize{$\mathcal{P}(\theta, t_2)$}}}
        \put(72,97){\tiny{Reference Scene}}
      \put(72,95.5){\tiny{Dynamic SFO}}   
     \put(72,93.8){\tiny{Static SFO} }

         \put(0.5,66.5){\rotatebox{90}{\scriptsize{$y$~(m)}}}
            \put(7,79.5){\scriptsize{\shortstack{Reference Scattered Fields, $t=t_0$\\ Re$\{E_z(x,y)\}$}}}
           \put(22,56.6){\scriptsize{\color{white}\textsc{PEC Wall}}}
            \put(22.4,72){\scriptsize{\color{white}\textsc{\shortstack{Dielectric \\Square} }}}
            \put(24,67.4){\scriptsize{\color{white}\textsc{\shortstack{PEC \\Object}}}}
            \put(7,75){\scriptsize{\color{white}\textsc{Observer}}}
         
        \put(31,66.5){\rotatebox{90}{\scriptsize{$y$~(m)}}}
             \put(38,79.5){\scriptsize{\shortstack{Reference Scattered Fields, $t=t_1$\\ Re$\{E_z(x,y)\}$}}}
            \put(53,56.6){\scriptsize{\color{white}\textsc{PEC Wall}}}
        
        \put(62,66.5){\rotatebox{90}{\scriptsize{$y$~(m)}}}
         \put(69,79.5){\scriptsize{\shortstack{Reference Scattered Fields, $t=t_2$\\ Re$\{E_z(x,y)\}$}}}
         \put(84,56.6){\scriptsize{\color{white}\textsc{PEC Wall}}}
    
         \put(16,20.5){\scriptsize{$x$~(m)}}
    \put(0.5,34){\rotatebox{90}{\scriptsize{$y$~(m)}}}
         \put(8,49){\scriptsize{\shortstack{Metasurface Scattered Fields, $t=t_0$\\ Re$\{E_z(x,y)\}$}}}
    
         \put(46,20.5){\scriptsize{$x$~(m)}}
    \put(31,34){\rotatebox{90}{\scriptsize{$y$~(m)}}}
         \put(38,49){\scriptsize{\shortstack{Metasurface Scattered Fields, $t=t_1$\\ Re$\{E_z(x,y)\}$}}}
    
         \put(77.5,20.5){\scriptsize{$x$~(m)}}
    \put(62,34){\rotatebox{90}{\scriptsize{$y$~(m)}}}
     \put(69,49){\scriptsize{\shortstack{Metasurface Scattered Fields, $t=t_2$\\ Re$\{E_z(x,y)\}$}}}

    \put(16,3){\scriptsize{$\theta$~(deg)}}
   \put(2,8){\rotatebox{90}{\scriptsize{Re\{$\chi_\text{ee}(\theta, t_0)$\}}}}
   	\put(21,44){\scriptsize{\color{white}\textsc{\shortstack{Motion\\Trajectory}}}}
    
    \put(47,3){\scriptsize{$\theta$~(deg)}}
    \put(32,8){\rotatebox{90}{\scriptsize{Re\{$\chi_\text{ee}(\theta, t_1)$\}}}}
    	\put(51,25){\scriptsize{\color{white}\textsc{\shortstack{Metasurface\\Hologram}}}}
    
    \put(78,3){\scriptsize{$\theta$~(deg)}}
   \put(63,8){\rotatebox{90}{\scriptsize{Re\{$\chi_\text{ee}(\theta, t_2)$\}}}}

    \end{overpic}
    \caption{A dynamic hologram with reconfigurable surface susceptibilities projecting an illusion of a specified object (PEC triangle, here) located within a slowly time-varying complex scene consisting of a PEC background and a moving dielectric square object along a specified trajectory.  The reference scattered fields, and the recreated fields using a metasurface hologram, the corresponding surface susceptibilities and the SFO scene, at three arbitrary time instants, a) $t_0$, b) $t_1$ and c) $t_2$. Illumination fields are the same as the incident fields. The simulation parameters are: dielectric inclusion of size $0.02 \times 0.02$~m with refractive index 2, PEC triangle of size 0.025~m located at the origin $(0,0)$ and the observer is located at $(-0.0525,0.0525)$~m.}
    \label{Fig:Dynamic}
\end{figure*}

\subsection{Dynamic Illusions}

So far, all the scenes were considered to be static with no time-variation. In many practical scenarios, the scene may be changing with time due to spatial movement of objects or addition or removal of some scene elements. In such cases, to maintain the illusion of a specified object, the metasurface hologram must reconfigure itself in response to the changes in the scene. There have been several works in the literature on devising novel reconfigurable metasurfaces across the electromagnetic spectrum where the basic unit cell building the metasurface, is either loaded with external tuning elements like PIN and varactor diodes, common at microwave frequencies or using exotic and advanced materials whose electronic properties may be tuned using external controls \cite{Reconfigr_MS,Reconfigr_MS2, Optical_MS_Reconfig}. While majority of the works on reconfigurable metasurfaces features slow time-variation, so that the modulation frequency is very small compared to the operation frequency of the surface, i.e. $\omega_p \ll \omega_0$, there has been a several research activities in the area of space-time modulation, where the modulations are comparable to operation frequencies of the surface \cite{STmod}. Let us consider the scenario of a slowly changing environment to illustrate the case of a dynamic illusion assuming reconfigurable surface susceptibilities of the metasurface hologram.

\begin{figure*}[htbp]
    \centering
                	 \begin{overpic}[grid=hide, scale = 0.27]{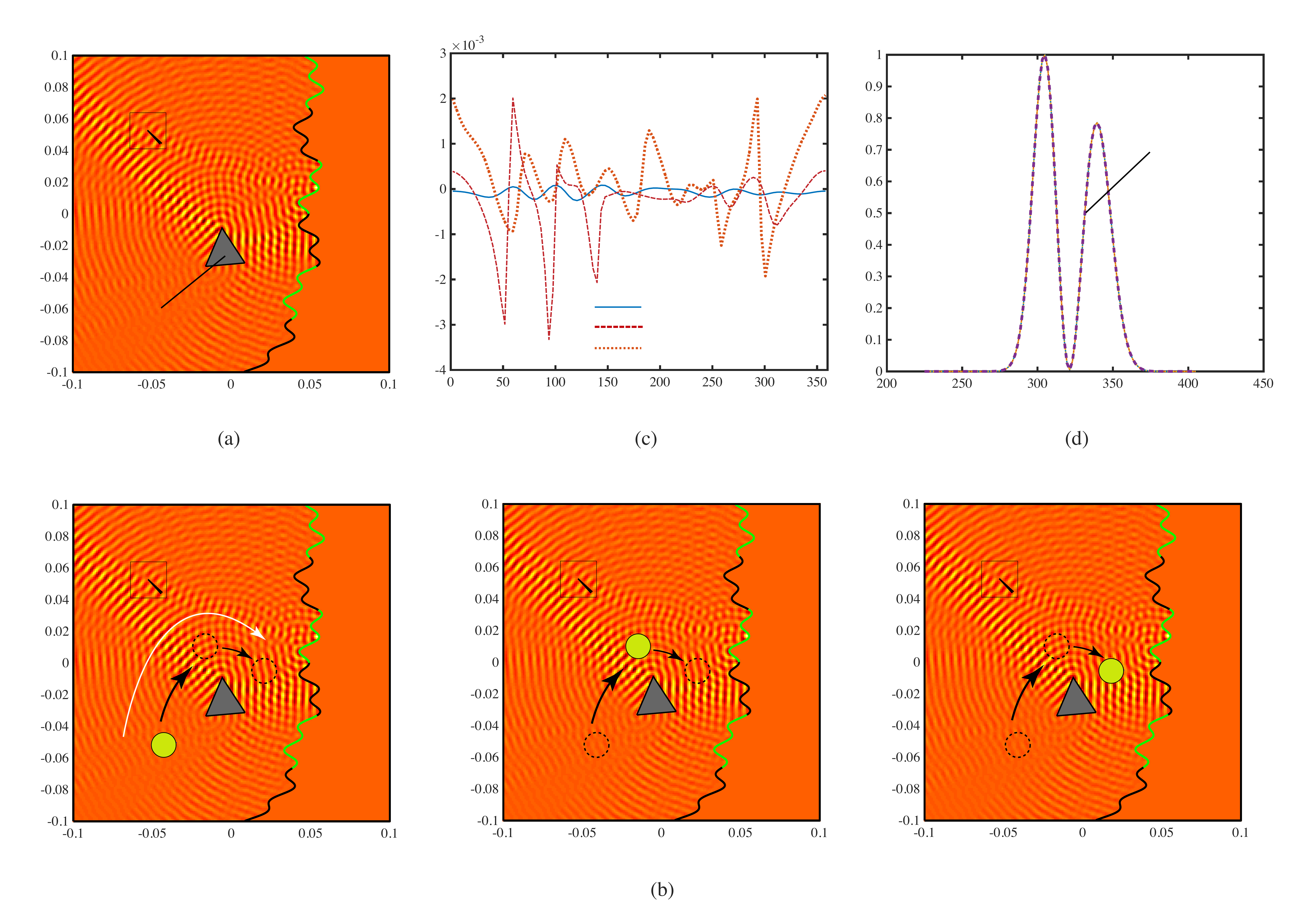} 
\put(16,4.5){\scriptsize$x$~(m)}
\put(16,38.5){\scriptsize$x$~(m)}
\put(49,4.5){\scriptsize$x$~(m)}
\put(81,4.5){\scriptsize$x$~(m)}
\put(1.5, 52){\scriptsize\rotatebox{90}{$y$~(m)}}
\put(1.5, 18){\scriptsize\rotatebox{90}{$y$~(m)}}
\put(34.5, 18){\scriptsize\rotatebox{90}{$y$~(m)}}
\put(66, 18){\scriptsize\rotatebox{90}{$y$~(m)}}
\put(47,38.5){\scriptsize$\theta$~(deg)}
\put(80,38.5){\scriptsize$\theta$~(deg)}
\put(31, 50){\scriptsize\rotatebox{90}{Re\{$\chi_\text{ee}(\theta)$\}}}
\put(64, 52){\scriptsize\rotatebox{90}{$\mathcal{P}(\theta)$}}
\put(22.5,43){\scriptsize\color{white}\textsc{\shortstack{Patterned \\Wall}}  }
\put(9,43){\scriptsize\color{white}\textsc{\shortstack{PEC \\Object}}}
\put(8,62){\scriptsize\color{white}\textsc{Observer}}
\put(86,59){\scriptsize\shortstack{ Reference Scene \\(\& $\forall t$)} }
\put(8,28){\scriptsize\color{white}\textsc{Observer}}
\put(41,28){\scriptsize\color{white}\textsc{Observer}}
\put(73,28){\scriptsize\color{white}\textsc{Observer}}
\put(8,32){\scriptsize\shortstack{Metasurface Scattered Fields, $t=t_0$\\ Re$\{E_z(x,y)\}$}}
\put(40,32){\scriptsize\shortstack{Metasurface Scattered Fields, $t=t_1$\\ Re$\{E_z(x,y)\}$}}
\put(72,32){\scriptsize\shortstack{Metasurface Scattered Fields, $t=t_2$\\ Re$\{E_z(x,y)\}$}}
\put(8,66.5){\scriptsize\shortstack{Reference Scattered Fields, $t=t_0$\\ Re$\{E_z(x,y)\}$}}
\put(13,24){\scriptsize\color{white}\textsc{\shortstack{Motion\\ Trajectory}}}
\put(45,23){\scriptsize\color{white}\textsc{\shortstack{Motion\\ Trajectory}}}
\put(9,9){\scriptsize\color{white}\textsc{\shortstack{Moving\\ Hologram}}}
\put(49,46.25){\tiny$t_0$}
\put(49,44.75){\tiny$t_1$}     
\put(49,43){\tiny$t_2$} 
    \end{overpic}
    \caption{Dynamic camouflaging example using reconfigurable metasurface susceptibilities, where a moving hollow hologram (hiding an arbitrary object) moves through a complex scene without being detected by the observer, and effectively camouflaging it. a) Reference scattered fields without the metasurface hologram. b) Computed scattered fields in the presence of a moving hologram at three times instants, and c) the corresponding time-dependent surface susceptibilities. d) The SFO scene detected by the observer at all time instants and the reference scene. Illumination fields are the same as the incident fields. The simulation parameters are: PEC triangle of side 0.025~m located at $(-0.005,-02)$~m, radius of the camouflaged region $ 0.0075$~m and the observer located at $(-0.0525,0.0525)$.}
    \label{Fig:Cloak}
\end{figure*}

Consider the reference scene of Fig.~\ref{Fig:Dynamic}(a) to be recreated. It consists of a square dielectric inclusion (DI) and a PEC triangular object in front of a perfectly opaque background. They are excited from the left using a Gaussian beam, and an observer located in the scene measures the scattered fields to image the scene. The DI moves through region on parameterized trajectory to simulate a changing environment, while the PEC object and the background remain fixed with respect to the observer. At time instant $t_0$, the scattered fields are computed and used to synthesize a metasurface with $\chi_\text{ee}(\theta, t_0)$. The PEC object is then removed and replaced by a circular metasurface hologram, successfully recreating the fields, as seen from the field and the SFO plots shown in Fig.~\ref{Fig:Dynamic}(a).

Next, the DI is allowed to move along a specified trajectory, which perturbs the originals scattered fields and the new reference scene is shown in Fig.~\ref{Fig:Dynamic}(b). The scattered fields are recomputed and a new metasurface is synthesized to recreate the scene at $t = t_1$, resulting in $\chi_\text{ee}(\theta, t_1)$. Consequently, the observer measures the new scene and deduce only the motion of the DI while falsely perceiving the presence the PEC object as before. Had the metasurface susceptibilities been constant, the scene would have changed as a result of the DI motion, which is clearly observed in the static SFO plot of Fig.~\ref{Fig:Dynamic}(b). This would have enabled the observer to detect a different and a distorted non-triangular PEC object, thereby breaking the illusion. This process of computing the scattered fields and synthesizing the metasurface hologram is repeated periodically, as long as the DI is in motion, as further illustrated for a third time instant $t=t_2$ in Fig.~\ref{Fig:Dynamic}(c)\footnote{A continuous animation of this process is provided in the supplementary information as a separate multimedia file.}.

Such a dynamic illusion naturally rests on the reconfigurable capability of the metasurface, and the time-scale at which such changes happen on the device level \cite{Nature_Cloak_AI}. An important step in this case is the sensing and prediction of the changes in the environment by either the metasurface itself or the software control unit configuring the surface, so that the scattered fields may be efficiently computed to synthesize the surface susceptibilities. The total time to reconfigure a metasurface hologram thus essentially depends on this intermediate step, where the actual reconfiguration of the surface is usually fast, typically in the order of few microseconds or less \cite{Reconfigr_MS,Reconfigr_MS2, Optical_MS_Reconfig}.

\subsection{Electromagnetic Camouflage}

As illustrated in Sec.~II, while electromagnetic camouflage is a sub-set of a general holograms, where the holograms projects its background to the observer to deceive it, it is an important practical application on its own. Thus let us consider an example of electromagnetically camouflaging a region in a complex scene. Fig.~\ref{Fig:Cloak}(a) shows a reference scene consisting of triangular PEC object placed in front of a patterned background. The background is modeled using a metasurface with alternating reflective and absorbing region to generate complex scattering fields. An incident Gaussian beam from the left is next applied which generate the scattered fields from the PEC object and the pattern background, in the entire region of interest. The observer simply detects the object and the background. Now, we wish to have an arbitrary object move and navigate through this region without being detected by the observer, thereby effectively camouflaging it at all times.

To achieve this, a metasurface skin could enclose the object of interest, and with internal illumination to mimic its composite background consisting of the PEC object and the patterned wall. Knowing the desired reference fields of Fig.~\ref{Fig:Cloak}(a), the required surface susceptibilities of the metasurface are computed and applied to the region. Fig.~\ref{Fig:Cloak}(a) shows the scene with the region included, and as expected, the reference fields are accurately reproduced in the entire region. Since, the observer measures the same fields, it does not detect the camouflaged region. As the region moves inside the scene, the metasurface susceptibilities must be recomputed at each time similar to that in Fig.~\ref{Fig:Dynamic}. Fig.~\ref{Fig:Cloak}(b) shows three time instants corresponding to three different positions of the camouflaged region along with slightly different spatial distributions of the surface susceptibilities shown in Fig.~\ref{Fig:Cloak}(c). In all cases, the SFO measured by the observer is the same, as shown in Fig.~\ref{Fig:Cloak}(d), and thus the camouflaged region remains undetected at all times\footnote{A continuous animation of this camouflaging process is provided in the supplementary information as a separate multimedia file.}.

\section{Conclusions}

A systematic numerical framework based on IE-GSTCs has been presented in 2D to synthesize closed metasurface holograms and skins for creating electromagnetic illusions of specified objects. The general hologram surface has been modeled using a zero-thickness sheet model of a generalized metasurface expressed in terms of its surface susceptibilities, which has been further integrated into the GSTCs. In combination with the IE current-field propagation operators, a simple yet powerful numerical framework of IE-GSTC has been developed. It has been further shown that the phenomenon of electromagnetic camouflage can be considered a special case of a general hologram principle, where instead of projecting the illusion of a specific object, the object wrapped in a metasurface skin may simply mimic its background by projecting that to the observer as the illusion. To estimate the effectiveness of the illusions, the notion of a scene constructed by an observer is developed from first principles and a simple mathematical model, termed as SFO, based on spatial Fourier transform has been proposed to construct the angular scene measured by an observer. The SFO method has been found to be more insightful and computationally superior to an other-wise brute-force simulation of the scene with a real detector. Next, using numerical examples, it has been shown that to recreate the reference desired fields everywhere in space using a closed metasurface hologram/skin, an internal illumination must be applied inside the hologram, in addition to the applied external illumination fields. Finally, several numerical examples have been presented of simple, angle-dependent and dynamic illusions, along with one example showing a dynamic camouflaged region of space, where it can freely move inside a given complex scene without being detected by the observer.

This work represents a continuation of the metasurface hologram synthesis of \cite{smy2020surface} focussing on open metasurfaces, which made it possible to have a clear delineation of the two half-spaces resulting in a simple classification of Front/back-lit Posterior/Anterior Illusion with Front/Back-illumination. The Hologram synthesis has been now extended to closed metasurfaces where the previous classification doesn't apply. Furthermore a requirement of two kinds of illuminations (external and internal) become a must for field recreation instead of a single front or back illumination of an open surface. Although implemented in 2D, the extension of the method to 3D is a straightforward task. A wide variety of 3D electromagnetic BEM code has been reported in the literature with even free software codes~\cite{BEMpp,PUMA-EMS,Puma-EM}, for instance. The integration of the GSTC interface equations with wide variety of BEM methods poses no fundamental issues and scaling the method to large 3D problems is simply a matter of implementation. Finally, the number of possible configurations and situations for creating EM illusions using both open and closed metasurface holograms are virtually unlimited, and so are the number of surface susceptibility solutions. While the few selected examples chosen here have been to highlight and illustrate various features of the hologram synthesis, the proposed framework represents a flexible test-bed to explore a wider variety of illusion scenarios useful for hologram designers before attempting practical demonstrations. Naturally, the unprecedented capabilities of EM metasurfaces in achieving very complex wave transformations is the core of this procedure, and this presented work in conjunction with \cite{smy2020surface}, represents a comprehensive set of tools to design and implement versatile metasurface based holographic surfaces and camouflaging skins.

\section*{Acknowledgements}

The authors acknowledge funding from the Department of National Defence's Innovation for Defence Excellence and Security (IDEaS) Program in support of this work.

\bibliographystyle{IEEEtran}
\bibliography{2020_Metasurface_Closed_Illusions_TAP_Smy}

\EOD

\end{document}